\documentclass[aps,prl,superscriptaddress,longbibliography,twocolumn]{revtex4-2}
\usepackage{epsf,graphicx}
\usepackage{amssymb}
\usepackage{amsmath}
\usepackage{latexsym,bm,array,amsfonts,multirow}
\usepackage{color}
\usepackage{ulem}
\usepackage{epstopdf}
\usepackage{orcidlink}
\usepackage{comment}
\newcommand{\ve}[1]{\boldsymbol{#1}}

\begin{document}
 \title{ Revealing altermagnetic 
 Fermi surfaces with two Kondo impurities }

	\author{Qiong Qin}
	\affiliation{New Cornerstone Science Laboratory, Department of Physics, School of Science, Westlake University, Hangzhou 310024, Zhejiang, China}

\author{Toshihiro Sato}
\affiliation{Institute for Theoretical Solid State Physics, IFW Dresden, 01069 Dresden, Germany}
\affiliation{Würzburg-Dresden Cluster of Excellence ctd.qmat, Germany}

\author{Marcin Raczkowski \orcidlink{0000-0003-1248-2695}}
\affiliation{Institut für Theoretische Physik und Astrophysik, Universität Würzburg, 97074 Würzburg, Germany}

\author{Jeroen van den Brink}
\affiliation{Institute for Theoretical Solid State Physics, IFW Dresden, 01069 Dresden, Germany}
\affiliation{Würzburg-Dresden Cluster of Excellence ctd.qmat, Germany}

\author{Congjun Wu}
\email[]{wucongjun@westlake.edu.cn}
\affiliation{New Cornerstone Science Laboratory, Department of Physics, School of Science, Westlake University, Hangzhou 310024, Zhejiang, China}
\affiliation{Institute for Theoretical Sciences, Westlake University, Hangzhou 310024, Zhejiang, China}
\affiliation{Key Laboratory for Quantum Materials of Zhejiang Province, School of Science, Westlake University, Hangzhou 310024, Zhejiang, China}
\affiliation{Institute of Natural Sciences, Westlake Institute for Advanced Study, Hangzhou 310024, Zhejiang, China}
\author{Fakher F. Assaad \orcidlink{0000-0002-3302-9243}}
\email[]{fakher.assaad@physik.uni-wuerzburg.de}
\affiliation{Institut für Theoretische Physik und Astrophysik, Universität Würzburg, 97074 Würzburg, Germany}
\affiliation{Würzburg-Dresden Cluster of Excellence ctd.qmat, Germany}

\begin{abstract}
Motivated by recent advances in the study of altermagnetism, or unconventional magnetism, and in the realization and manipulation
of  two-impurity Kondo 
physics in real materials, we propose a phase-sensitive method to explore
unconventional magnetic symmetries.
Our method can be implemented with
spin-resolved scanning tunneling microscopy to study  two-impurity Kondo phenomena on altermagnetic metals by varying the distance and orientation between 
magnetic impurities.
Using 
quantum Monte Carlo simulations, 
we analyze the spin splitting of the Kondo resonance,  
whose spatial distribution 
sensitively captures the symmetry of the underlying altermagnetic order. 
Furthermore, 
the impurity spin correlations reflects
the anisotropy of the RKKY interaction
due to the altermagnetic Fermi surface
splitting. 
This work provides a framework for studying the competition between the Kondo effect, the RKKY interaction and altermagnetism, in the simplest possible system.    
\end{abstract}
\maketitle

\textit{Introduction.--} 
Unconventional magnetism (UM) \cite{Wu2004,Wu2007a} was proposed based on Fermi surface instability in analogy to unconventional superconductivity two decades ago. 
It has recently attracted extensive attention \cite{Naka:2019aa,Ahn2019,Hayami2019,Smejkal2022,Bai2023,Ouassou2023,Yaqian,Lee2024,Liu2024d,Liu2024e,Ding2024,Amin2024,McClarty2024,Lu2024a,Das2024,Ghorashi2024,Sato2024,Liao2024,Zhou2024,Jin2024,Chen2024PRX,Jiang2024PRX,Xiao2024PRX,Yang2025,Jiang2025,Song2025,Gu2025,Chen2025,Consoli2025,Lin2025,Duan2025,Pupim2025,Sato025}, driven by experimental realizations and its potential applications in spintronics--particularly due to its absence of stray magnetic fields and suitability for  terahertz (THZ) high-frequency operations. 
Most existing studies have focused on its origin, classification, and interplay with superconductivity. 
However, its interaction with other quantum phenomena, such as Kondo physics, remains largely unexplored \cite{Zhao2025,Diniz2024,Lee2025}. 
This raises a compelling question: could novel and intriguing physics emerge from the interplay between Kondo effects and unconventional magnetism?

To detect unconventional magnetism -- particularly 
altermagnetism (AM) -- transport measurements and spin-resolved angle-resolved photoemission spectroscopy (ARPES) have been extensively employed. 
In the case of ARPES \cite{Lee2024,Liu2024d,Ding2024,Smejkal2022,Yang2025,Jiang2025,
Song2025}, the identification primarily relies on observing spin splitting in single-particle spectra and comparing the experimental Fermi surfaces with those predicted by Density functional theory (DFT) calculations. 
Additionally, signatures such as the anomalous Hall effect and other transport phenomena \cite{Song2025} have been used as indirect evidence for the presence of AM. 
As in the theory of superconductivity, \textit{smoking gun} experiments are phase-sensitive and directly probe the relative phase of an order parameter  \cite{Tsuei00}.  Our motivation is to propose a scanning-tunneling-microscopy (STM) experiment
in the realm of Kondo physics with a spin polarized tip \cite{Wiesendanger09}.

\begin{figure}[b]
\begin{center}
\includegraphics[width=8cm]{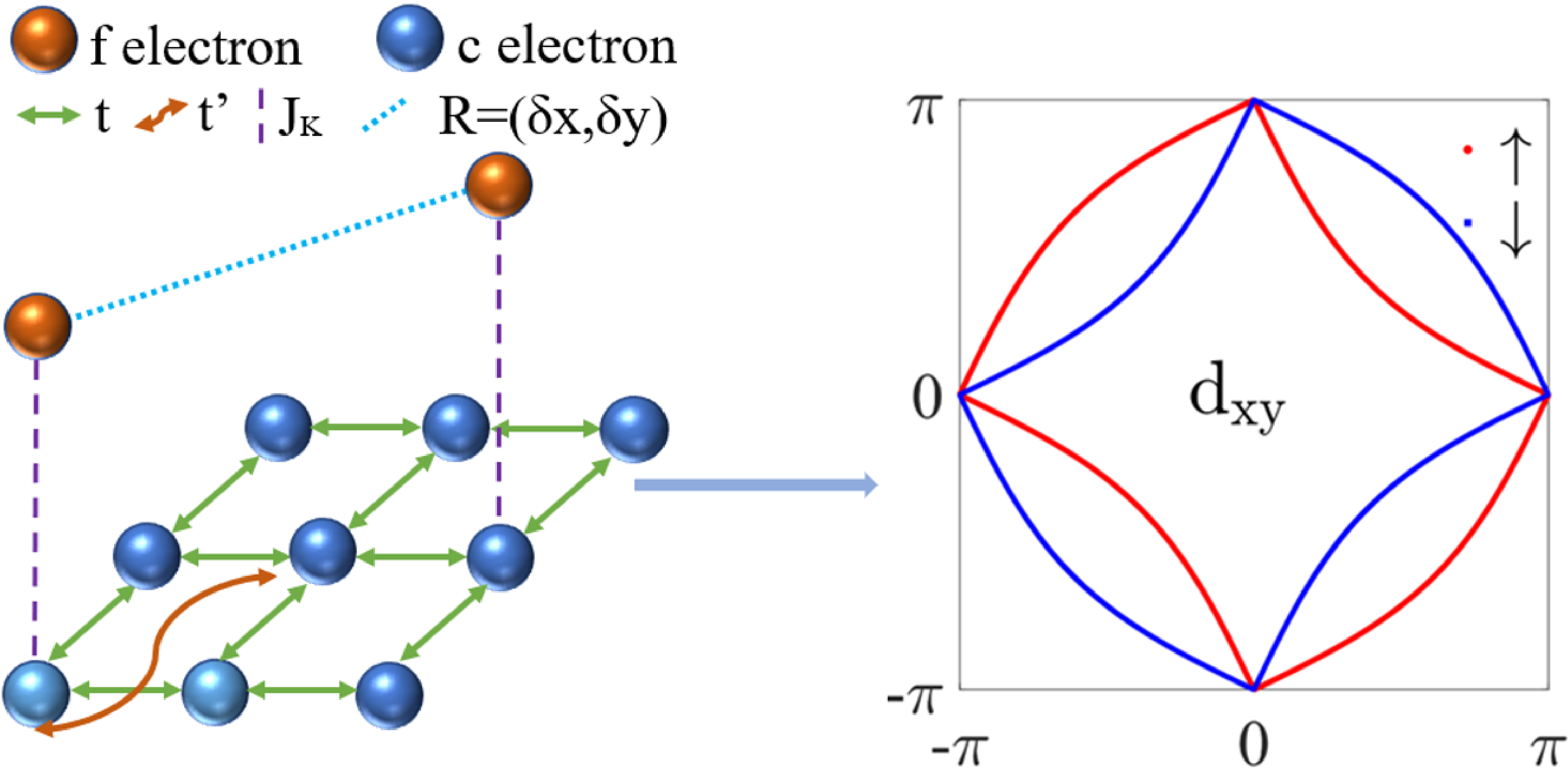}
\end{center}
\caption{
Illustration of the two-impurity Kondo model, with the right panel depicting the Fermi surface of the underlying AM metal with $t'=0.4$.
}
\label{fig1}
\end{figure}

Motivated by the above questions and experimental needs, we propose a phase-sensitive approach based on Kondo physics to detect AM (a form of unconventional collinear magnetism). 
Specifically, we incorporate an  AM term into the conduction electron sector of the two-impurity Kondo model \cite{Jayaprakash1981,Jones1988,Fye1989,Sire1993,He2015,Dong2021b,Spinelli2015,VonBergmann2015,Campo2004}, as schematically illustrated in Fig.~\ref{fig1}, and perform 
unbiased finite-temperature quantum Monte Carlo (FTQMC) simulations \cite{Blankenbecler81,White89,Assaad08_rev}.
Our results show that Kondo resonance can sensitively reflect the underlying symmetry of the AM order, clearly distinguishing between two AM states of $d_{x^2 - y^2}$ and $d_{xy}$ symmetries. 
These symmetry features are also manifested in the spin-spin correlation functions mediated by the RKKY interaction. 
Moreover, we observe a crossover between the two-impurity and single-impurity regimes as the inter-impurity distance varies. 
The temperature-dependence analysis further reveals the competition between the RKKY interaction, Kondo effect and AM. 
Based on these findings, we propose that spin-resolved (SR)-STM measurements on magnetic impurities embedded in an AM metal can serve as a powerful phase-sensitive probe 
for detecting AM.

\textit{Model and method.--}
We consider the two-impurity Kondo model Hamiltonian 
 with the conduction electrons exhibiting the $d$-wave-type AM band structure,
\begin{equation}
    H = \sum_{\bm{k},\sigma} \epsilon_{\bm{k}\sigma} \hat{c}^{\dagger}_{\bm{k}\sigma} \hat{c}^{\phantom\dagger}_{\bm{k}\sigma} + J_{K}\sum_{j=1,2}\bm{\hat{S}}_j\cdot\bm{\hat{s}}_j.
\label{eq:Kondo}
\end{equation}
$J_K$ describes the on-site antiferromagnetic Kondo coupling between localized magnetic impurities and conduction electrons with $\bm{\hat{S}}_j$ representing the  
impurity spin with $s=\frac{1}{2}$ at site $j$; $\bm{\hat{s}}_j$ is the spin 
operator of conducting electrons on site $j$, which is defined as $\bm{\hat{s}}_j = \sum_{\alpha\beta} \hat{c}_{j\alpha}^\dagger \frac{\bm{\sigma}_{\alpha\beta}}{2} \hat{c}_{j\beta}$ and $\hat{c}_{j\sigma}^\dagger$ creates a conduction electron at site $j$ with the $z$-component of spin $\sigma$.
The  dispersion relation is given by: 
$\epsilon_{\bm{k}\sigma}= -2t [ \cos k_x + \cos k_y ] - 2t' \text{sgn}(\sigma)
[\cos(k_x +  k_y) - \cos(k_x - k_y) ]$.
$t$ denotes the nearest-neighbor hopping and is set to unity
below and $t'$ is a next-nearest-neighbor hopping, yielding the $d_{xy}$-type spin splitting. Here, $\hat{c}_{\bm{k}\sigma}^\dagger$ creates a conduction electron  in  Bloch state  $\bm{k}$ and $z$-component of spin $\sigma$.

In the absence of impurities,  the Bloch  Hamiltonian possesses the spin group symmetry $R_{sg}$, {\it i.e.}, the spatial rotation at $\frac{\pi}{2}$ followed by the spin rotation at $\pi$ flipping the $z$-component of spin, the $C_{2v}$  with  mirror planes along the
diagonal directions.
We will also study the two-impurity Kondo problem with $d_{x^2-y^2}$-type AM band structure.   In this case, the dispersion relation reads: 
$\epsilon^{\prime}_{\bm{k}\sigma} = -2t [\cos k_x + \cos k_y ] - 2t' \text{sgn}(\sigma)
[\cos 2k_x - \cos 2k_y ]$.
The $d_{x^2-y^2}$  Hamiltonian has the same symmetry as the $d_{xy}$  provided that  the mirror reflection  planes of the $C_{2v}$ operations are along the $x$ or $y$ direction.

For the finite temperature auxiliary-field determinant QMC simulations, we  adopt 
an Abrikosov fermion representation of the impurity spin \cite{Coleman2015} as, 
$\bm{\hat{S}}_j = \sum_{\alpha\beta} \hat{f}_{j\alpha}^\dagger \frac{\bm{\sigma}_{\alpha\beta}}{2} \hat{f}_{j\beta}$, subject to the single-occupancy constraint $\sum_{\sigma} \hat{f}_{j\sigma}^\dagger \hat{f}_{j\sigma} = 1$ for $j = 1, 2$. 
The model enjoys the combined 
time-reversal (TR) and particle-hole symmetry as defined by  $\hat{c}^{\dagger}_{j,\sigma} \rightarrow \sum_{s} i\sigma^{y}_{\sigma,s} e^{i\ve{Q}\cdot\ve{j}}  \hat{c}^{\phantom\dagger}_{js}$ with $\ve{Q} =(\pi,\pi)$.
This symmetry property of the Hamiltonian allows for a negative sign free  FTQMC formulation  of  the model of Eq.~(\ref{eq:Kondo}) for both 
AM symmetries \cite{Wu04,SatoT17_1}. 
Our implementation follows Refs.~\cite{Assaad99a,Capponi00} and we have used the Algorithms for Lattice Fermions  (ALF) \cite{ALF_v2.4} implementation of the  FTQMC. 
The lattice size is taken as $L=12$ for the simulations.

\textit{Numerical results.--}
To at best understand how the  AM band  impacts the Kondo effect, it is convenient to formulate the action in terms of impurity spin $\ve{n}_{i}(\tau)$ and fermion $\ve{c}^{\dagger}_{i}(\tau)$ coherent states. 
One can then integrate out the fermionic degrees of freedom that do not couple to 
the impurity spin so as to obtain the action: 
\begin{eqnarray}
    \label{Eq:Action_Kondo}
    S  &=& S_{0} +   J_{K} \sum_{i =1}^{2} \int_{0}^{\beta} d\tau  s \ve{n}_{\ve{i}}(\tau) \cdot \ve{c}^{\dagger}_{i}(\tau) (\frac{\ve{\sigma}}{2}) \ve{c}^{\phantom\dagger}_{i}(\tau)   \\ 
&  + &
\int_{0}^{\beta}\int_{0}^{\beta}d \tau  d \tau' \sum_{i,j=1}^{2}  \ve{c}^{\dagger}_{i}(\tau) \ve{G}_{\bm{0}}^{-1}(\tau-\tau',i-j) \ve{c}^{\phantom\dagger}_{j}(\tau'),\nonumber
\end{eqnarray}
where $\left[\ve{G}_{\bm{0}}(\tau-\tau',i-j') \right]_{\sigma,\sigma'} $ $ =   \langle T_{\tau} \hat{c}^{\dagger}_{j,\sigma'}(\tau') \hat{c}^{\phantom\dagger}_{i,\sigma}(\tau)  \rangle_{0}$  is 
 the conduction electron Green function between the  impurity sites in the absence of the Kondo coupling,  
and $S_{0}$  is the Berry phase of the spin-1/2 degrees of freedom, $s \ve{n}_i(\tau)$ with 
$\ve{n}_i$ the unit vector on the $S^2$ sphere.   
The  AM symmetries of the band are encoded in the spin dependence 
of the single particle Green function.  As shown  in the End Matter,  
if the displacement vector $\mathbf{R}=(\delta x, \delta y)$ between sites $1$ and $2$ is parallel to the nodal line in momentum space,   
then the Green function 
shows no spin  dependence: 
It is proportional to the unit matrix in spin space. 
As a consequence, for this special 
orientation
SU(2) spin symmetry  is  restored and we expect no spin splitting of the Kondo resonance.  This equally implies that  single impurity problems with or without the AM term are symmetry equivalent.  
If the two impurities are not on a nodal line, then the Green  function will acquire a spin dependence and we will  see that this  leads to a spin splitting of the Kondo resonance.

To capture the Kondo resonance in the realm of the Kondo model, we consider the composite fermion operator:
$\hat{\psi}_{j\alpha}^{\dagger} = (\hat{c}_{j}^{\dagger} \bm{\sigma})_{\alpha} \cdot \hat{\bm{S}}_j$ \cite{Costi00,Danu2021,Raczkowski2022}.  This quantity transforms as the electron creation operator  
and the associated imaginary-time Green function is defined as, $\tilde{G}_{j\sigma}(\tau)=\langle T_{\tau} \hat{\psi}_{j\sigma}(\tau)\hat{\psi}_{j\sigma}^{\dagger}(0)\rangle$. 
The spin-resolved and frequency-dependent spectral function $\tilde{N}_{\sigma}(\omega)$ is obtained using the maximum entropy method  via the spectral decomposition at $\tau>0$,
\cite{Jarrell1996}:
\begin{eqnarray}
\tilde{G}_{j,\sigma}(\tau)=\int d\omega \frac{\tilde{N}_{j,\sigma}(\omega)e^{-\omega\tau}}{e^{-\beta\omega}+1}.
\end{eqnarray}

\begin{figure}[t]
	\begin{center}
		\includegraphics[width=8cm]{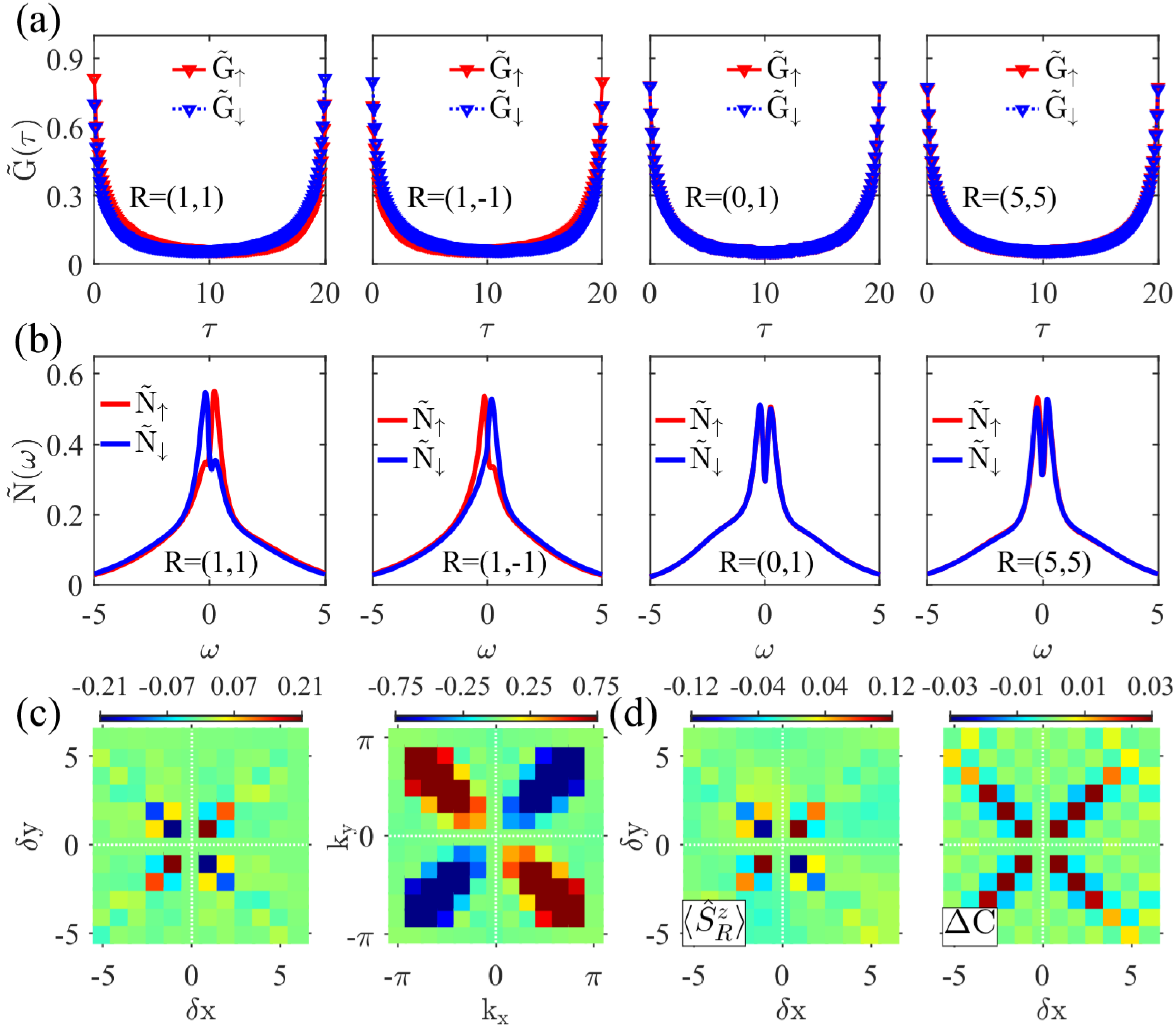}
	\end{center}
	\caption{ $d_{xy}$ AM symmetry at $J_{K} = 2$ and $t^{\prime}=0.4$.
		(a) Imaginary-time Green function of the composite fermion, $\tilde{G}_{\bm{R}}(\tau)$, for varying impurity separations $\bm{R} = (\delta x, \delta y)$, where $\bm{R}$ denotes the distance between the two magnetic impurities.
		(b) Corresponding spectral function $\tilde{N}_{\ve{R}}(\omega)$ for different $\bm{R}$.
		(c) Real-space distribution of the difference $\Delta \tilde{G}_{\ve{R}}$ between spin-up and spin-down Green functions, together with its Fourier transform.
		(d) Local magnetic moment $\langle \hat{S}_{\bm{R}}^z\rangle$ and the real-space equal-time spin correlation functions between the two impurities, showing the anisotropy $\Delta C_{\bm{R}} = C^{x}_{\bm{R}}(0^+) - C^{z}_{\bm{R}}(0^+)$.
		Temperatures are $T = 0.05$ for (a)(b), and $T = 0.10$ for (c)(d).
		}
	\label{fig2}
\end{figure}  

In Fig.~\ref{fig2} (a) and (b), the Green function $\tilde{G}_{\ve{R},\sigma}(\tau)$ and the corresponding spectral function $\tilde{N}_{\ve{R},\sigma}(\omega)$ exhibit a clear spin-dependent asymmetry in the presence of $d$-wave AM splitting, {\it i.e.}, $t^\prime \neq 0$. 
When $t' = 0$, $\tilde{N}_{\ve{R},\sigma}(\omega)$ is spin-degenerate.
At finite values of $t'$, the peak asymmetries with respect to $\pm \omega$ become spin-dependent, and are opposite for $\tilde{N}_{\ve{R},\uparrow}(\omega)$ and $\tilde{N}_{\ve{R},\downarrow}(\omega)$, respectively. 

The spin-dependence of the asymmetry in $\tilde{N}_{\ve{R},\sigma}(\omega)$ relies on the orientation of 
the displacement vector
$\bm{R}$. 
It becomes strongest when $\bm{R}$ is along the anti-nodal  directions, 
say,
$(1,1)$, for the $d_{xy}$ type of AM.
If $\bm{R}$ is rotated by $\frac{\pi}{2}$, the spin-dependence is reversed, consistent with the spin-group symmetry $R_{sg}$.
When
$\bm{R}$ is along the nodal direction, $(0,1)$  for $d_{xy}$ AM, the total Hamiltonian possesses the combined symmetry of $R_{sg}$ followed by real-space  reflection with respect 
anti-nodal line.  
This symmetry protects the degeneracy of 
the Green functions and spectral functions: $\tilde{G}_{\ve{R},\uparrow}(\tau) =\tilde{G}_{\ve{R},\downarrow}(\tau)$ and consequently, $\tilde{N}_{\ve{R},\uparrow}(\omega) = \tilde{N}_{\ve{R},\downarrow}(\omega)$.

To further characterize the spatial distribution of the spin asymmetry, we define the integrated difference in the composite fermion Green function as
\begin{eqnarray}
	\Delta \tilde{G}_{\ve{R}}&=&\left(\int_{0}^{\beta/2}d\tau-\int_{\beta/2}^{\beta}d\tau\right)\left(\tilde{G}_{\ve{R},\uparrow}(\tau)-\tilde{G}_{\ve{R},\downarrow}(\tau)\right)\nonumber\\
    &=&\int d\omega \left[ \tilde{N}_{ \ve{R},\uparrow}(\omega)-\tilde{N}_{\ve{R},\downarrow}(\omega)\right] f_{\beta}(\omega), 
\label{eq:DeltaG}
\end{eqnarray}
with $f_{\beta}(\omega) = \frac{1}{\omega} \left( 1 - \frac{1}{\cosh(\beta \omega/2)}\right)$ an odd function of frequency.In contrast, the sum of integrations from $0$ to $\beta/2$ and from $\beta/2$ to $\beta$ is just
the difference between 
$\tilde{G}_{\ve{R},\uparrow}(\omega=0)$ and
$\tilde{G}_{\ve{R},\downarrow}(\omega=0)$, yielding 0 due to the symmetry $\tilde{N}_{\ve{R},\uparrow}(\omega)=
\tilde{N}_{\ve{R},\downarrow}(-\omega)$. $\Delta \tilde{G}_{\ve{R}}$
vanishes  in the presence of TR symmetry \footnote{This includes the generic two impurity problem on a metallic surface where spin-orbit coupling is present},  and acts as a proxy for the asymmetry between up and down local density of  states.  
A key point is that this quantity can be computed from the imaginary-time Green function without having to use analytical continuation. 
The dependence of $\Delta \tilde{G}$ on $\bm{R}$ 
is shown in Fig.~\ref{fig2}(c) along with its Fourier transformation in terms of  $(k_x, k_y)$.  Here, we carry out a simulation for each value of $\ve{R}$  and then carry out the Fourier transformation.
We see that $\Delta \tilde{G}_{\ve{R}}$ changes sign under a $\pi/2$ rotation, 
resulting into nodal lines along the directions
$\delta x =0$ and $\delta y=0$, and along $k_x,~k_y=0$ in its Fourier transform, which directly reflects the underlying $d_{xy}$ symmetry of the AM 
 background.

Due to the breaking of TR  symmetry, static magnetic moments appear in the presence of two impurities. 
The rotation of $\pi$ around the $z$-axis with respect to the middle point of two impurities exchanges them, hence, the 
moments on the impurity sites should be the same.
In Fig.~\ref{fig2}(d), the value of $\langle \hat{S}^{z}_{\ve{R}} \rangle$ 
is depicted as a function of $\ve{R}$.
The AM symmetry of the band is  also captured by the anisotropy in the spin fluctuations between two impurities.  Consider,
\begin{eqnarray}
	C_{\bm{R}}^{x(z)}(\tau)=\left\langle T_{\tau}\hat{S}_{\bm{0}}^{x(z)}(\tau) \hat{S}_{\bm{R}}^{x(z)}\right\rangle -
     \left\langle\hat{S}_{\bm{0}}^{x(z)}\right\rangle \left\langle\hat{S}_{\bm{R}}^{x(z)}\right\rangle,
\end{eqnarray}
then we define the spin anisotropy as the difference between the equal time spin-spin correlation functions as
$\Delta C_{\ve{R}}=C_{\bm{R}}^x(0^+) - C_{\bm{R}}^z(0^+)$, which exhibits nodal behavior along the $x$ or $y$ directions, consistent with the $d_{xy}$-wave symmetry
as shown in Fig.~\ref{fig2}(d), where the equal-time convention is taken in the limit of $\tau\to 0^+$. 
This follows from Eq. ~(\ref{Eq:Action_Kondo})  
and the restoration of the SU(2) spin symmetry for impurities along a nodal line. 
In the End Matter similar computations are carried out  for the  $d_{x^2-y^2}$ AM symmetry. 

\begin{figure}[t]
\begin{center}
		\includegraphics[width=8cm]{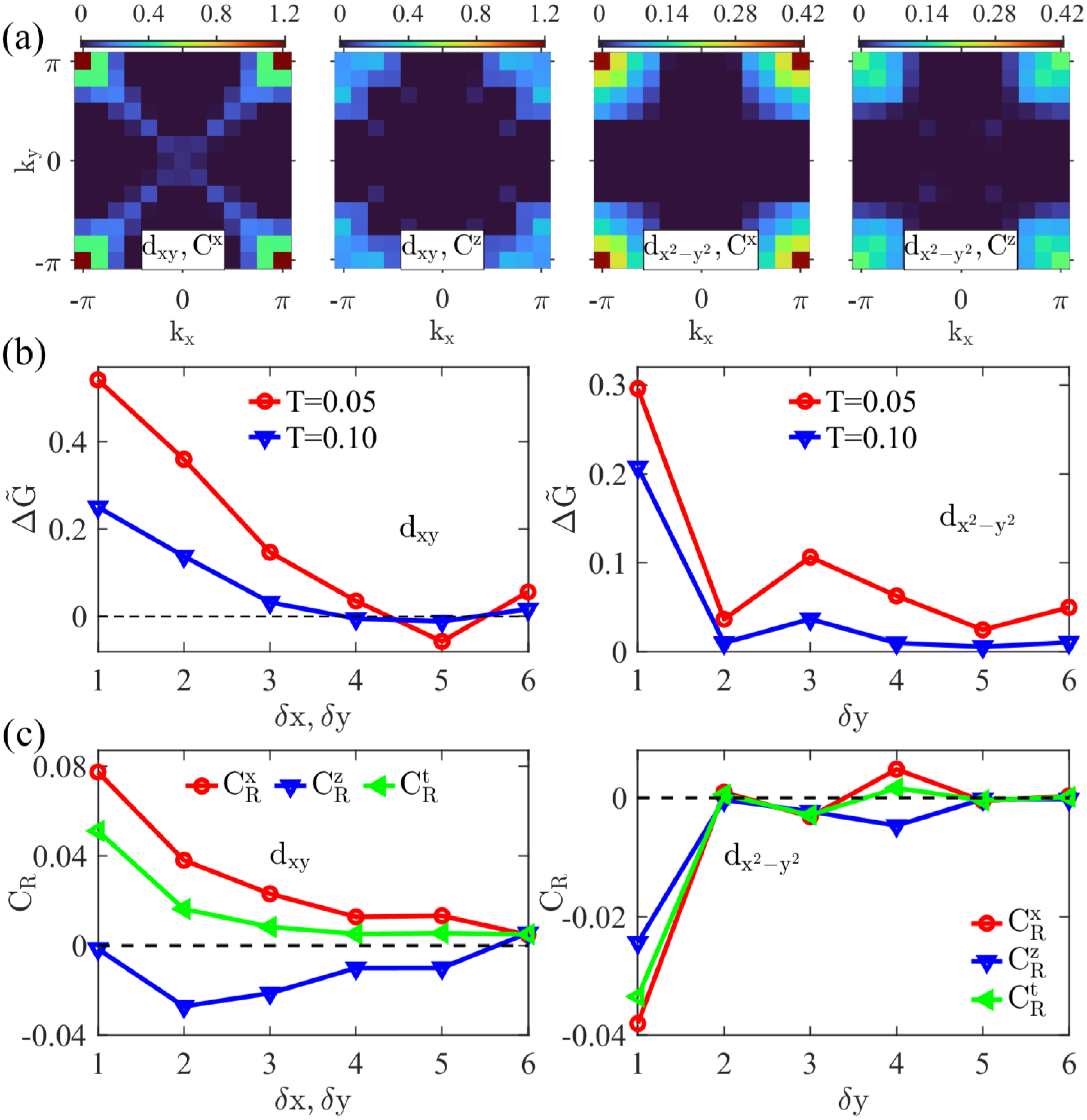}
	\end{center}
	\caption{The RKKY interaction at $J_{K}=2$ and $t^{\prime}=0.4$.
		(a) Spatial Fourier transform of the equal-time spin correlation functions $C^{x,z}(0^+)$ between two impurities for both the $d_{xy}$ and $d_{x^2 - y^2}$ AM cases.
		(b) Distance dependence of the composite fermion Green function difference $\Delta \tilde{G}_{\ve{R}}$ for the $d_{xy}$ and $d_{x^2 - y^2}$ cases at $T=0.05,~0.10$.
		(c) Distance dependence of the spin correlation $C_{\bm{R}}^{x,z}(0^+)$ between impurities and
        $C_{\bm{R}}^{t}(0^+)=\frac{1}{3}[2C_{\bm{R}}^{x}(0^+)+C_{\bm{R}}^{z}(0^+)]$.
        The direction analyzed for $d_{xy}$ case is along the diagonal $x = y$ direction, whereas for the $d_{x^2 - y^2}$ case it is along the $y$-axis. We note $T=0.1,~0.05$ for (a),~(c), respectively. 
}
	\label{fig3}
\end{figure}  

The AM term has important consequences on the nature of the RKKY interaction. 
Adopting a combined spin and fermion coherent state path integral formulation of the Kondo model, one can integrate out the fermionic degrees of freedom.   
The
second-order perturbation theory in the Kondo coupling yields the RKKY interaction that reads 
\begin{eqnarray}
S_{\rm RKKY} &=& \frac{J_K^2s^2}{2} \int_{0}^{\beta} d \tau  \int_{0}^{\beta} d \tau'
n_{\bm{0},\mu}(\tau) C^{0,\mu}_{\bm{R}}(\tau - \tau') \nonumber \\
&\times&  n_{\bm{R},\mu}(\tau'),
\end{eqnarray}
where $\ve{n}_{\mu}(\tau) $ is a spin coherent state, and $C^0_{\bm{R}}(\tau)$ is the conduction electron spin correlation tensor evaluated in the non-interacting limit.   
Owing to the U(1) spin symmetry of the AM band structure, it is diagonal and differs in its 
$xy$ and $z$ components, i.e.,
$C^{0,x}=C^{0,y}\neq C^{0,z}$.
Due to the nesting properties of the conduction electron Fermi surfaces for spin-up and spin-down components, the transverse susceptibility $\chi^{0,x}$  diverges logarithmically at the nesting vector $\bm{Q} = (\pi, \pi)$ as a function of  temperature. The longitudinal  susceptibility $\chi^{0,z}$ on the other hand shows no such divergence. 

In the interacting  QMC simulations, the spatial Fourier transform of $C_{\bm{R}}^{x}(0^+)$ 
exhibits a pronounced maximum at the wave vector $\bm{Q} = (\pi, \pi)$, and $C^{x}_{\bm{Q}}(0^+)$ is significantly larger than 
$C^{z}_{\bm{Q}}(0^+)$, as illustrated in Fig.~\ref{fig3}(a). 
At $T=0$, the RKKY  interaction  follows a power-law  in real space such that the effect  of one impurity on the other is not set by  
a characteristic length scale.   
Fig.~\ref{fig3}(b) depicts the spin splitting of the Kondo resonance,
$\Delta \tilde G_{\ve{R}}$,  as a function of $\ve{R}$. As the temperature  decreases, the range of the splitting increases for both AM symmetries.  As typical for a quantity without a characteristic length scale that fits on the considered finite site lattice, we observe strong size effects in terms of an upturn of the  magnitude of the splitting  at the largest distance, $\ve{R}=(6,6)$, of our $L=12$ lattice.   

The  anisotropy of the  spin  correlations along the anti-nodal directions is illustrated in Fig.~\ref{fig3}(c).   
While in the  
$x$-components we observe (anti)-ferromagnetic 
correlation between sites on (different)  same sub-lattices,  the  $z$-component  shows dominant anti-ferromagnetic correlations.  The transverse spin  correlations
stand in agreement   with  the expectations  of the RKKY interaction in the absence of the AM term. On the other hand,  the behavior of the  longitudinal component stems from the AM band structure.

\begin{figure}[t]
	\begin{center}
		\includegraphics[width=8cm]{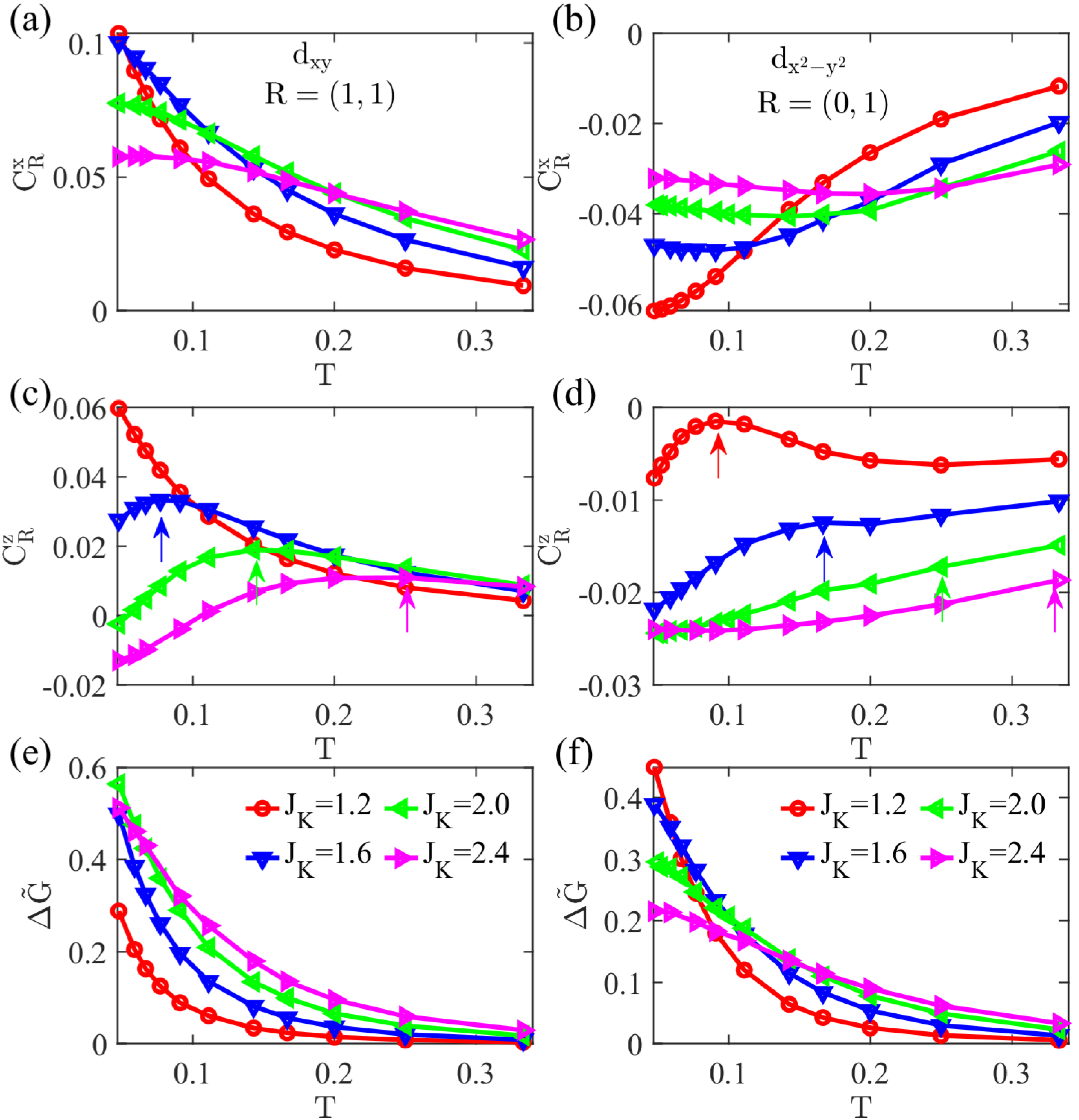}
	\end{center}
	\caption{Temperature evolutions for the RKKY (a, b, c,d) and for Kondo splitting (e, f) at $t^{\prime} = 0.4$ for $J_K = 1.2,~1.6,~2.0$.
		(a,b) The $x$-component equal-time spin correlation function $C_{\bm{R}}^{x}(0^{+})$ as a function of temperature for the $d_{xy}$ and $d_{x^2 - y^2}$ cases.
		(c,d) Spin correlation along $z$ direction $C_{\bm{R}}^{z}(0^{+})$ as a function of temperature for the $d_{xy}$ and $d_{x^2 - y^2}$ symmetry. The arrows 
        denote the $T_\chi$ scale, see  main text and End Matter. 
		(e,f) Temperature dependence of the composite fermion Green function difference $\Delta \tilde{G}_{\ve{R}}$ at $\bm{R} = (1,1)$ for the $d_{xy}$ case and at $\bm{R} = (0,1)$ for the $d_{x^2 - y^2}$ case.
}
	\label{fig4}
\end{figure}

Figure~\ref{fig4}  plots the  temperature dependence of spin correlations and spin splitting of the Kondo resonance  for both  AM symmetries and demonstrates the subtle interplay between scales of the problem. 
One relevant scale  is $T_\chi$ below which the local impurity spin susceptibility tends to saturate.  When the distance between the two impurities is such that they effectively decouple,  this quantity  maps onto the  single impurity Kondo scale.  The determination of this quantity is  detailed in the End Matter,  and  the results are plotted as arrows in Figs.~\ref{fig4}(c)(d).  Another scale  is  the  RKKY interaction that can compete \cite{Doniach77} or coexist  \cite{Capponi00,Danu2021} with Kondo screening. Finally, the AM  term  feeds into the RKKY interaction, and  provides the spin-splitting of the Kondo resonance.  
Below $T_\chi$ and in the large $J_K$ limit where the RKKY interaction can be omitted corresponds to the domain of validity of the large-$N$ mean-field calculations presented in
the Supplemental Material.  

For the $d_{x^2-y^2}$ symmetry, see Figs.~\ref{fig4}(b)(d)(f),  we can resolve  $T_\chi$  for  all  considered  values of $J_K$.   This stems from the fact that the density of  states of the AM band structure  shows a van Hove  singularity at the Fermi energy. 
In the  low temperature  regime  $T/T_{\chi} \ll 1 $  we observe i) the  anisotropy between the transverse, Fig.~\ref{fig4}(b), and longitudinal Fig.~\ref{fig4}(d), spin correlations is the biggest at small  $J_K$  where the RKKY interaction  dominates over the  Kondo scale \cite{Doniach77} and ii) the  spin-splitting of the Kondo resonance 
decreases as a function of growing values of $J_K$. This behavior follows from the observation that the  effective  magnetic field perceived by the impurity spin decreases  when measured in units of  $T_{\chi}$.  On the other hand $T/T_{\chi} > 1$ we note that i)  $\Delta \tilde{G}$ still takes a finite value and grows as a function of  $J_K$ ii) the spin  anisotropy is much smaller and dominated by the AM band structure. 

For $d_{xy}$ symmetry, the van Hove singularity at the Fermi energy is absent and the  $T_\chi$ and RKKY scales are smaller in the small $J_K$ limit. In fact, at $J_K=1.2, 1.6$  the presented data in Figs.~\ref{fig4}(a)(c)(d) are dominated by the high temperature behavior discussed above,  whereas $J_K= 2, 2.4$ capture  high and  low temperature behaviors.

\textit{Discussion.--}
In spin-resolved STM measurements \cite{Spinelli2015,VonBergmann2015}, 
the Kondo-related features can be easily identified in the differential conductance spectra. 
By extracting the peak heights, one can define a spin splitting parameter $\eta=\frac{h_{\uparrow}-h_{\downarrow}}{h_{\uparrow}+h_{\downarrow}}$, where $h_{\uparrow(\downarrow)}$ are the resonance peak values for spin-up and -down polarizations, respectively, located at either the positive or the negative bias.  
The real-space distribution of $\eta$, along with its Fourier transform, may directly reveal the symmetry characteristics of the underlying AM order. 
Such spatially resolved measurements can be achieved by varying the location and orientation of the second magnetic impurity. 
Furthermore, complementary techniques sensitive to local magnetization, such as muon-spin-rotation ($\mu$SR), could also be employed to detect the symmetry of the AM phase.

The spectral features resemble those seen in the two-impurity Kondo problem subject to an external magnetic field. 
Therefore, our setup, realized in the absence of an  applied magnetic field, provides a platform to explore rich 
Kondo phenomena of two impurities, 
including the singlet and triplet Kondo states\cite{Campo2004,Spinelli2015}, 
two-stage Kondo screening \cite{Jayaprakash1981}, 
as well as single-impurity Kondo behavior.

\textit{Conclusion.--}
In summary, we have performed finite-temperature QMC simulations on the two-impurity Kondo model incorporating AM terms in the conduction band. 
The real-space spin splitting of the Kondo resonance reveals the symmetry associated with the AM terms. 
Our results provide a  phase-sensitive approach for detecting AM via spin-resolved STM measurements.  The two impurity Kondo system is the  simplest possible model to understand the competition and intertwinement between Kondo, RKKY  and  AM.   Further work  for  denser 
systems is left for  future investigations.  

\textit{Acknowledgments.--}  The authors thank Kun Yang, Danqing Hu, Gaopei Pan and Mengfan Wang for helpful discussions.
QQ and CW are supported by National Natural Science Foundation of China (Grants No. 12447125, No. 12234016) and the New Cornerstone Science Foundation. They also acknowledge the computation resource provided by Westlake HPC Center.
TS and MR acknowledge 
financial support by the Deutsche Forschungsgemeinschaft (DFG, German Research Foundation) through the Würzburg-Dresden Cluster of Excellence ctd.qmat – Complexity, Topology and Dynamics in Quantum Matter (EXC 2147, project-id 390858490).
FFA thanks financial support through the
AS 120/16-2 (Project number 493886309) that is part of the collaborative research project SFB QMS funded by the Austrian Science Fund (FWF) F 86.

\bibliography{refs.bib}

\onecolumngrid
\clearpage
\appendix*
\begin{center}
{\bf\large End Matter}
\end{center}
\twocolumngrid
\setcounter{equation}{0}

\textit{Symmetries of the single particle Green function.--}
Here we  investigate the spin symmetry of the single particle Green function.  

%
\begin{figure}[tb]
\begin{center}
\includegraphics[width=8cm]{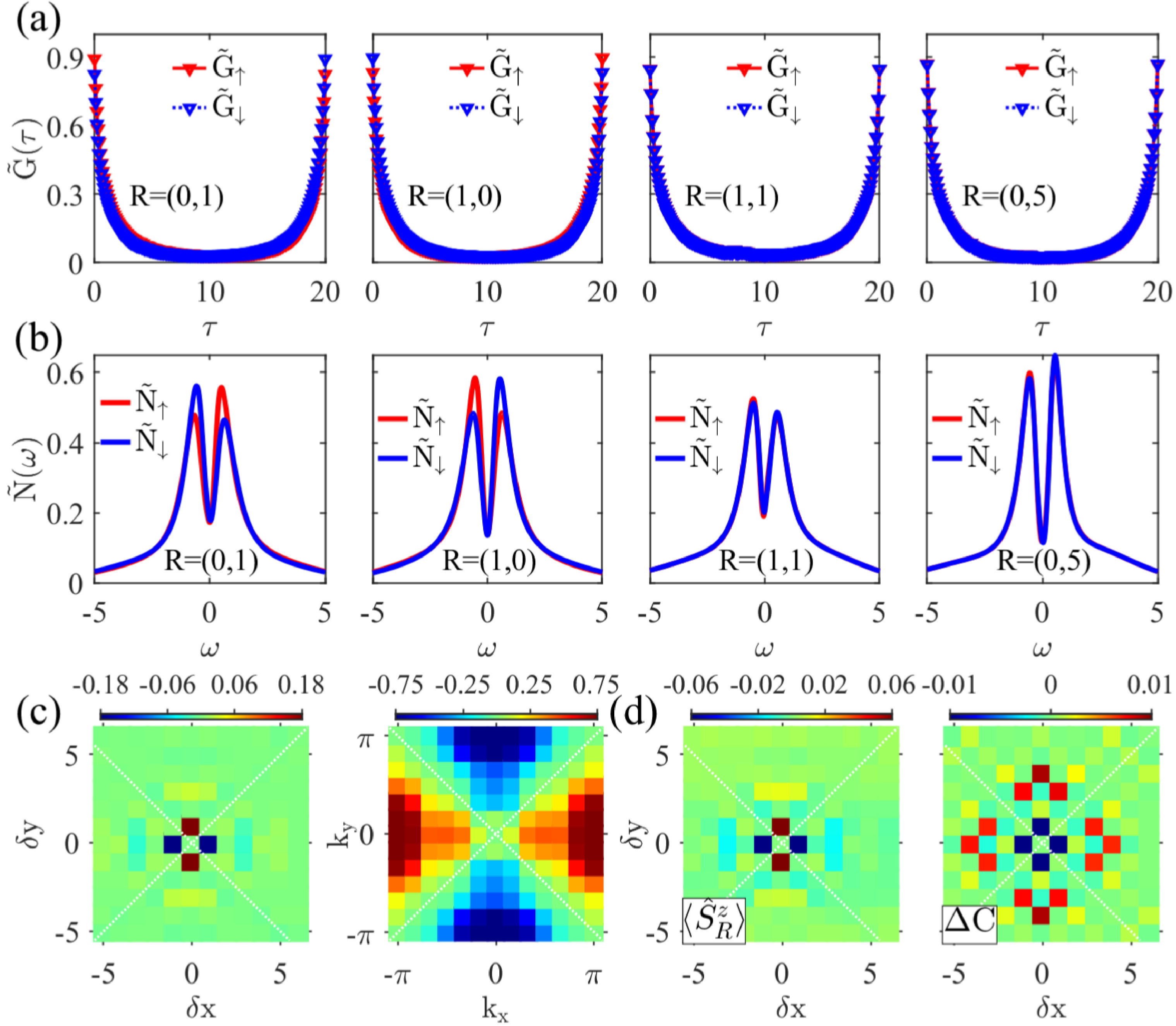}
\end{center}
\caption{$d_{x^2 - y^2}$ altermagnetic symmetry at $J_{K} = 2$ and $t^{\prime}=0.4$.
(a) Imaginary-time Green function of the composite fermion $\tilde{G}_{\ve{R}}(\tau)$ at different displacement vector $\bm{R} = (\delta x, \delta y)$ between two impurities.
(b) Spectral function $\tilde{N}_{\ve{R}}(\omega)$ at different values of $\bm{R}$.
(c) Real-space distribution of $\Delta \tilde{G}_{\mathbf{R}}$
defined in Eq. (\ref{eq:DeltaG}) along with its Fourier transform. 
(d)The local magnetic moment $\langle \hat{S}_{\bm{R}}^{z} \rangle$ and the real-space equal-time spin correlation functions between two impurities exhibit anisotropy, characterized by $\Delta C_{\bm{R}} = C^{x}_{\bm{R}}(0^+) - C^{z}_{\bm{R}}(0^+)$.
Temperatures are set to $T = 0.05$ for panels (a) and (b), and $T = 0.10$ for panels (c) and (d).
}
\label{fig5}
\end{figure}
We will consider the real space Green function,
\begin{equation}
  G_{\sigma\sigma^\prime}(i-j, \tau)   =    \langle T \hat{c}^{\dagger}_{i,\sigma^{\prime}}(\tau) \hat{c}^{\phantom\dagger}_{j,\sigma} (0) \rangle.
\end{equation}
and show that if $i$ and $j$ are on a nodal line,  then the Green function is spin independent.

Firstly,  let 
$\hat{U}(e_{x(y)},\pi)$ be an global SU(2) rotation   around the $x(y)$-axis   with angle $\pi$, and  $\hat{T}_m$ be  the mirror symmetry  on the nodal lines. That is $(1,1)$, $(1,-1)$  for  $d_{x^2 - y^2}$   and  $(1,0),(0,1)$  for  $d_{xy}$.  
We define combined transformations 
\begin{eqnarray}
    \hat{T}_x = \hat{U}(e_x,\pi) \hat{T}_m, 
     \, ~
     \hat{T}_y =\hat{U}(e_y,\pi) \hat{T}_m,
\end{eqnarray}
where
\begin{eqnarray}
\hat{T}_x \ve{\hat{c}}^{\dagger}_{i} \hat{T}^{-1}_x  &=&  \ve{\hat{c}}^{\dagger}_{T_m(i)} U(e_x,\pi),\\
\hat{T}_y \ve{\hat{c}}^{\dagger}_{i} \hat{T}^{-1}_y  &=&  \ve{\hat{c}}^{\dagger}_{T_m(i)} U(e_y,\pi), 
\end{eqnarray}
with $U(e_x,\pi) =  e^{i \pi \sigma_x/2} $ and
$U(e_y,\pi) =  e^{i \pi \sigma_y/2} $. 
Furthermore, one will show that:
\begin{equation}
  \left[\hat{H}_0, \hat{T}_x \right] = 
  \left[\hat{H}_0, \hat{T}_y \right]=0.
\end{equation}

Let us now assume that $i$ and $j$ both  
lie on a nodal line such that 
$T_m(i) = i$  and  $T_m(j) = j$.
We expand the $2\time 2$ matrix of $G(i-j,\tau)$
in terms of the identity and three Pauli matrices as
\begin{equation}
G(i-j,\tau)=a_0 I + a_i \sigma_i.
\end{equation}
The $\hat T_x$ symmetry ensures that $a_y=a_z=0$
and the $\hat T_y$ symmetry ensures that $a_x=a_z=0$,
hence, only $a_0\neq 0$ such that the  Green function along a nodal line has no spin dependence.

Generically, the Green function will be diagonal in the spin indices.  This follows from the U(1) spin-rotational symmetry about the $z$-axis that leaves  $\hat{H}_0$ invariant.

\textit{Real space spin splitting of the $d_{x^2-y^2}$ symmetry.--} The case of the $d_{x^2-y^2}$-type
altermagnetism can be analyzed within the same framework. 
The key distinction lies in the symmetry of the induced quantities: $\Delta \tilde{G}_{\ve{R}}$, $\langle\hat{S}_{\bm{R}}^{z}\rangle$ and $\Delta \chi_{\bm{R}}$ now exhibit nodal lines along the diagonal-direction, reflecting a characteristic $d_{x^2-y^2}$ symmetry, as shown in Fig.~\ref{fig5}. 

\begin{figure}[tbh]
	\begin{center}
		\includegraphics[width=8cm]{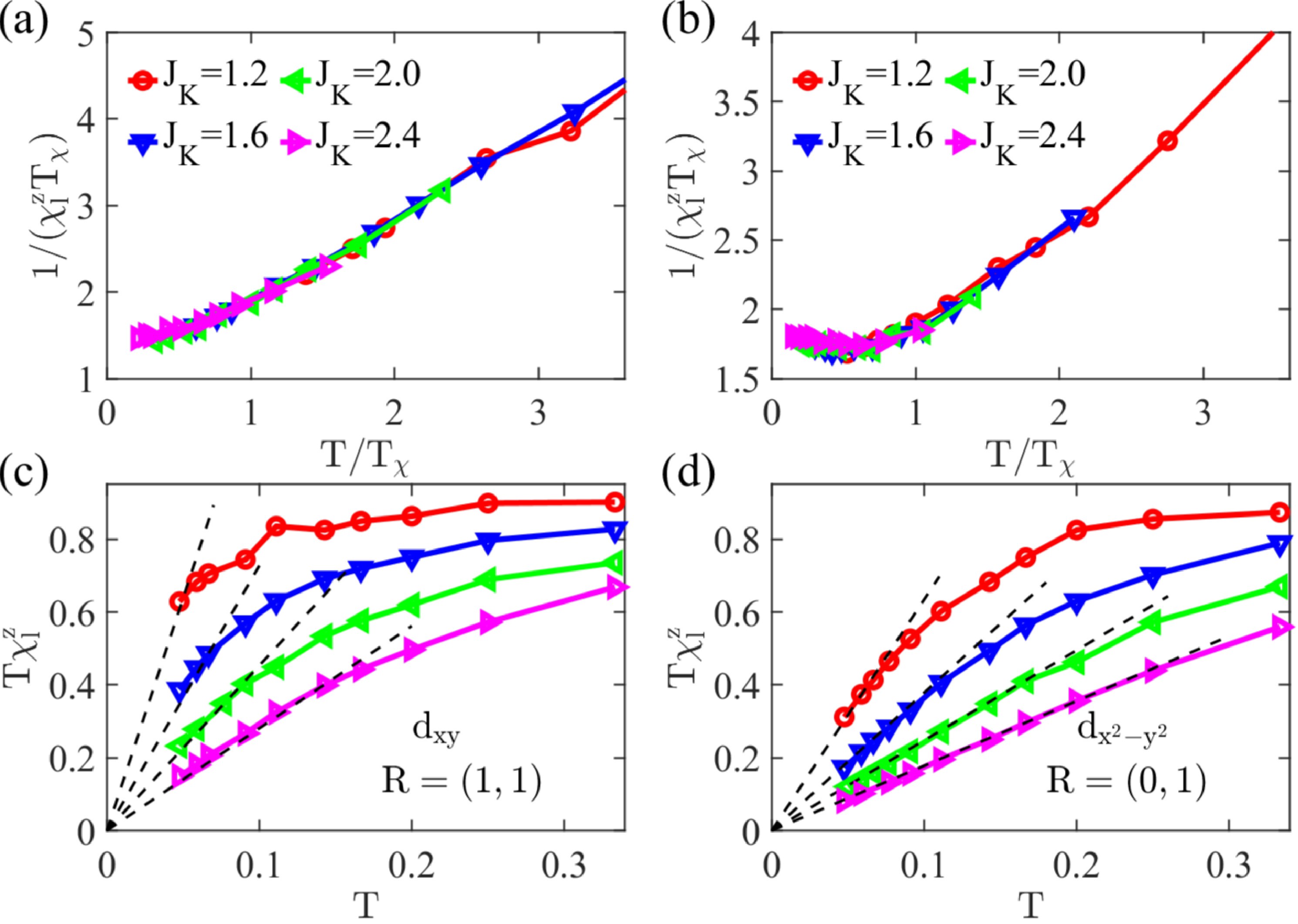}
	\end{center}
	\caption{ 
Local spin  susceptibility for $d_{xy}$ case (a)(c) with $\bm{R}=(1,1)$ and $d_{x^2-y^2}$ symmetry (b)(d) with $\bm{R}=(0,1)$. (a)(c) Temperature dependence of the local spin susceptibility $\chi_l^{z}$ for
$J_K = 1.2,\,1.6,\,2.0,\,2.4$, for $d_{xy}$ and $d_{x^2-y^2}$ respectively.
(b)(d) Temperature dependence of $T \chi_l^{z}$ for
$J_K = 1.2,\,1.6,\,2.0,\,2.4$, for $d_{xy}$ and $d_{x^2-y^2}$ respectively.
}
\label{fig6}
\end{figure}

\textit{Kondo screening.--}  We further analyze the effect of the altermagnetic term on Kondo screening. As mention in the main text, since the altermagnetic term acts as a momentum-dependent effective magnetic
field, rather than a uniform Zeeman field, it does not induce a breakdown of the single impurity 
Kondo effect.    For the two impurity case we have seen in the main text that impurities develop  a finite magnetization provided that they are not on a nodal line. This is very similar to the Kondo effect  in a magnetic field, and we will inspire ourselves from this piece of physics \cite{Andrei80,Hewson93,Costi00} to  interpret our data.   

We calculate the local spin correlation function
$C_l^{x(z)}(\tau) = \langle T_{\tau} \hat{S}_{\bm{0}}^{x(z)}(\tau) \hat{S}_{\bm{0}}^{x(z)} \rangle-\langle \hat{S}_{\bm{0}}^{x(z)}\rangle^2$
and its imaginary-time integral,
$\chi_l^{x(z)} = \int_0^\beta d\tau \, C_l^{x(z)}(\tau)$,
to characterize  Kondo screening. 
The latter quantity  corresponds to the local spin susceptibility.

As shown in Fig.~\ref{fig6}(a), the data for $\chi_l^{z}$ seemingly collapses onto a single curve
when rescaled by an appropriate characteristic temperature $T_{\chi}$ for
$J_K = 1.2,\,1.6,\,2.0,\,2.4$ for the $d_{xy}$ case with $\bm{R} = (1,1)$. Similar results are 
plotted in Fig.~\ref{fig6}(b)  for the $d_{x^2-y^2}$ case at  $\bm{R} = (0,1)$. 
As mentioned above, we would like to interpret these results in the framework of the single impurity  Kondo model in a magnetic field. Here,
the local spin susceptibility   satisfies  the scaling  relation
$T_K \chi_l^{z} =  f(T/T_K, B/T_K)$  with $f$  a scaling function \cite{Hewson93,Andrei80}.  When the  two impurities are very far apart, this form certainly holds since the impurities decouple. Furthermore, in this limit the effective magnetic field vanishes, leading to a data collapse of $T_K \chi_l^{z}$  versus $T/T_K$. As  the distance between the two impurities is reduced we  expect to observe correction to scaling when plotting $T_K \chi_l^{z}$  versus $T/T_K$  at constant magnetic field (i.e. constant value of $t'$).  The RKKY interaction is equally expected to yield corrections to scaling.  As apparent from the data in  Figs.~\ref{fig6}(a)(b) such correction to scaling seem to be small, and/or may be included in a redefinition of the Kondo scale, $T_\chi$.

The characteristic temperature $T_{\chi}$ can be identified from the deviation
from linear behavior for $T\chi_l^{z}$ shown in  Figs.~\ref{fig6}(c)(d). Furthermore, $\lim_{T \to 0} T \chi_l^{z} \to 0$,
indicating the absence of a local magnetic moment at zero temperature.  

Together, these results demonstrate that Kondo screening persists even in the
presence of altermagnetic terms. In the Supplemental Material, we show that the transverse component $\chi_l^{x}$ exhibits
nearly identical behavior, with only a negligible differences compared to
$\chi_l^{z}$. 

The  $T_{\chi}$  scale in the small $J_K$ limit turns out to be much smaller  for the  $d_{xy}$ case  as compared to the $d_{x^2-y^2}$ symmetry. 
We  understand this in terms of the local density of states:  for the $d_{x^2-y^2}$ ($d_{xy}$) symmetry we do (not) observe a Van-Hove singularity at 
the Fermi energy.  In particular, and as apparent from Fig.~\ref{fig6}(c) it becomes hard to pin down a value of $T_{\chi}$ within our accessible  
temperature  range and lattice size  for the $d_{xy}$ case at $J_K \lesssim 1.6$.

\onecolumngrid

\newpage
\onecolumngrid
\begin{appendix}
	\newpage
		\begin{center} {\large \textbf{Revealing altermagnetic Fermi surfaces with two Kondo impurities \\
Supplemental Material}}
\end{center}
		\renewcommand{\thefigure}{S\arabic{figure}}
		\renewcommand{\theequation}{S\arabic{equation}}
		\setcounter{equation}{0}
		\setcounter{figure}{0}
		
\setcounter{table}{0}
\section{I. Quantum Monte Carlo Simulations of the Multi-Impurity Kondo Model}

The simulations of  the SU($N$) multi-impurity Kondo model are carried out   along the lines of the Algorithms for Lattice fermions \cite{ALF_v2.4} implementation of the Kondo lattice model.  The Hamiltonian can be written as
\begin{eqnarray}
	H=H_{c}+H_{f}+H_{cf},
\end{eqnarray}
where the individual terms are given by
\begin{eqnarray}
	H_c&=&-t\sum_{i}\left(\hat{c}_{i\sigma}^{\dagger}\hat{c}_{i+\hat{x}\sigma}+\hat{c}_{i\sigma}^{\dagger}\hat{c}_{i+\hat{y}\sigma}+h.c.\right)-t^{\prime}\sum_{i\sigma}{\rm sgn}(\sigma)\left(\hat{c}_{i\sigma}^{\dagger}\hat{c}_{i+2\hat{x}\sigma}-\hat{c}_{i\sigma}^{\dagger}\hat{c}_{i+2\hat{y}\sigma}+h.c\right),\nonumber\\
	H_f&=&\frac{U_f}{N}\sum_{l}(\hat{n}_{l}^{f}-\frac{N}{2})^2,~~~~H_{cf}=\frac{2J_{K}}{N}\sum_{l}\sum_{a=1}^{N^2-1}\hat{T}_{l}^{a,c}\hat{T}_{l}^{a,f},
\end{eqnarray}
The operators appearing in the interaction term are defined as
\begin{eqnarray}
	\hat{T}_{l}^{a,c}=\sum_{\sigma,\sigma'}\hat{c}_{l\sigma}^{\dagger}T_{\sigma\sigma'}^a\hat{c}_{l\sigma'},~~~~~\hat{T}_{l}^{a,f}=\sum_{\sigma,\sigma'}\hat{f}_{l\sigma}^{\dagger}T_{\sigma\sigma'}^a\hat{f}_{l\sigma'},~~~~~\hat{n}_{l}^f=\sum_{\sigma=1}^{N}\hat{f}_{l\sigma}^{\dagger}\hat{f}_{l\sigma},~~~~\hat{n}_{l}^c=\sum_{\sigma=1}^{N}\hat{c}_{l\sigma}^{\dagger}\hat{c}_{l\sigma}.
\end{eqnarray}

The hybridization term can be recast in the form
\begin{eqnarray}
	H_{cf}=-\frac{J_{K}}{4N}\sum_{l}\left[(\hat{D}_l^{\dagger}+\hat{D}_l)^2+(i\hat{D}_l^{\dagger}-i\hat{D}_l)^2\right]+\frac{J_{K}}{4}n_{imp},
\end{eqnarray}
where we have employed the completeness relation
\begin{eqnarray}
\sum_{a}\hat{T}_{\alpha\beta}^a\hat{T}_{\alpha'\beta'}^a=\frac{1}{2}\left(\delta_{\alpha\beta'}\delta_{\alpha'\beta}-\frac{1}{N}\delta_{\alpha\beta}\delta_{\alpha'\beta'}\right),
\end{eqnarray}
and introduced the operators
\begin{eqnarray}
\hat{D}_l^{\dagger}=\sum_{\sigma=1}^{N}\hat{c}_{l\sigma}^{\dagger}\hat{f}_{l\sigma},~~~~\hat{D}_l=\sum_{\sigma=1}^{N}\hat{f}_{l\sigma}^{\dagger}\hat{c}_{l\sigma}.
\end{eqnarray}

For the SU(2) case, the QMC-related Hamiltonian takes the form
\begin{eqnarray}
	H=H_c+H_f+H_{cf},
\end{eqnarray}
with the individual contributions given by
\begin{eqnarray}
	H_c&=&-t\sum_{i}\left(\hat{c}_{i\sigma}^{\dagger}\hat{c}_{i+\hat{x}\sigma}+\hat{c}_{i\sigma}^{\dagger}\hat{c}_{i+\hat{y}\sigma}+h.c.\right)-t^{\prime}\sum_{i\sigma}{\rm sgn}(\sigma)\left(\hat{c}_{i\sigma}^{\dagger}\hat{c}_{i+2\hat{x}\sigma}-\hat{c}_{i\sigma}^{\dagger}\hat{c}_{i+2\hat{y}\sigma}+h.c\right),\nonumber\\
	H_f&=&\frac{U_f}{2}\sum_{l}(\hat{f}_{l\uparrow}^{\dagger}\hat{f}_{l\uparrow}+\hat{f}_{l\downarrow}^{\dagger}\hat{f}_{l\downarrow}-1)^2,~~~~~~~~~~H_{cf}=-\frac{J_{K}}{8}\sum_{l}\left[\sum_{\sigma}\left(\hat{c}_{l\sigma}^{\dagger}\hat{f}_{l\sigma}+\hat{f}_{l\sigma}^{\dagger}\hat{c}_{l\sigma}\right)\right]^2.\nonumber
\end{eqnarray}

After applying the Trotter decomposition, performing the discrete Hubbard–Stratonovich (HS) transformation, and integrating out the conduction electrons, the system becomes amenable to quantum Monte Carlo (QMC) simulations. Under the HS transformation, the interaction terms in the action are rewritten as
\begin{eqnarray}
S_f^{lk}=\frac{U_f\Delta \tau}{2}(f_{lk\uparrow}^{\dagger}f_{lk\uparrow}+f_{lk\downarrow}^{\dagger}f_{lk\downarrow}-1)^2\rightarrow \frac{U_f\Delta \tau}{2}\left[\lambda_{lk}^2-2i\lambda_{lk}(f_{lk\uparrow}^{\dagger}f_{lk\uparrow}+f_{lk\downarrow}^{\dagger}f_{lk\downarrow}-1)\right],
\end{eqnarray}
and 
\begin{eqnarray}
	S_{cf}^{lk}=-\frac{J_{K}\Delta \tau}{8}\left[\sum_{\sigma}\left(c_{lk\sigma}^{\dagger}f_{lk\sigma}+f_{lk\sigma}^{\dagger}c_{lk\sigma}\right)\right]^2\rightarrow \frac{J_K\Delta \tau}{8}\left[V_{lk}^2+2V_{lk}\sum_{\sigma}\left(c_{lk\sigma}^{\dagger}f_{lk\sigma}+f_{lk\sigma}^{\dagger}c_{lk\sigma}\right)\right].
\end{eqnarray}
Here, $l$ denotes the impurity site, $k$ labels the imaginary-time slice $\tau_k$, $\lambda_{lk}$ represents the onsite density field, and $V_{lk}$ denotes the onsite hybridization field.

We then employ the following approximation formula:
\begin{eqnarray}
	\int_{-\infty}^{\infty}dx e^{-x^2-2x\sqrt{\alpha}b}\approx\frac{\sqrt{\pi}}{4}\sum_{n=\pm1,\pm2}\gamma(n)e^{-\eta(n)\sqrt{\alpha}b},
\end{eqnarray}
where the coefficients are defined as
\begin{eqnarray}
	\gamma(\pm)=1+\frac{\sqrt{6}}{3},~~\gamma(\pm2)=1-\frac{\sqrt{6}}{3},~~\eta(\pm)=\pm\sqrt{2(3-\sqrt{6})},~~\eta(\pm2)=\pm\sqrt{2(3+\sqrt{6})}.
\end{eqnarray}
After this transformation, the auxiliary fields $\lambda_{lk}$ and $V_{lk}$ take only discrete values given by $\eta(\pm1,\pm2)$, with the corresponding discrete indices $n_{lk},~v_{lk}=\pm1,\pm2$. For the term $S_f^{lk}$, the parameters are identified as
\begin{eqnarray}
	\sqrt{\alpha}=f_{lk\uparrow}^{\dagger}f_{lk\uparrow}+f_{lk\downarrow}^{\dagger}f_{lk\downarrow}-1,~~b=i\sqrt{U_f\Delta\tau/2}.
\end{eqnarray}
Similarly, for the Kondo part $S_{cf}^{lk}$, 
\begin{eqnarray}
\sqrt{\alpha}=\sum_{\sigma}\left(c_{lk\sigma}^{\dagger}f_{lk\sigma}+f_{lk\sigma}^{\dagger}c_{lk\sigma}\right),~~b=\sqrt{J_{K}\Delta\tau/8}.
\end{eqnarray}

 Under the partial particle–hole transformation,
 \begin{eqnarray}
 	c_{lk\uparrow}\rightarrow c_{lk\uparrow},~~~~c_{lk\downarrow}\rightarrow (-1)^{l}c_{lk\downarrow}^{\dagger},~~~~~f_{lk\uparrow}\rightarrow f_{lk\uparrow},~~~~f_{lk\downarrow}\rightarrow -(-1)^{l}f_{lk\downarrow}^{\dagger},
 \end{eqnarray}
 the interaction terms transform as
 \begin{eqnarray}
 	i\sqrt{U_f\Delta\tau/2}\eta(n_{lk})(f_{lk\downarrow}^{\dagger}f_{lk\downarrow}-\frac{1}{2})&\rightarrow& -i\sqrt{U_f\Delta\tau/2}\eta(n_{lk})(f_{lk\downarrow}^{\dagger}f_{lk\downarrow}-\frac{1}{2}),\nonumber\\
 	\sqrt{J_{K}\Delta\tau/8}\eta(v_{lk})\left(c_{lk\downarrow}^{\dagger}f_{lk\downarrow}+f_{lk\downarrow}^{\dagger}c_{lk\downarrow}\right)&\rightarrow& \sqrt{J_{K}\Delta\tau/8}\eta(v_{lk})\left(c_{lk\downarrow}^{\dagger}f_{lk\downarrow}+f_{lk\downarrow}^{\dagger}c_{lk\downarrow}\right). 
 \end{eqnarray}
 Similarly, the kinetic terms transform as
 \begin{eqnarray}
 - t\sum_{i}\left(c_{ik\downarrow}^{\dagger}c_{i+\hat{x},k\downarrow}+c_{ik\downarrow}^{\dagger}c_{i+\hat{y},k\downarrow}+h.c.\right)\rightarrow- t\sum_{i}\left(c_{ik\downarrow}^{\dagger}c_{i+\hat{x},k\downarrow}+c_{ik\downarrow}^{\dagger}c_{i+\hat{y},k\downarrow}+h.c.\right)\nonumber\\
 t^{\prime}\sum_{i}\left(c_{ik\downarrow}^{\dagger}c_{i+2\hat{x},k\downarrow}-c_{ik\downarrow}^{\dagger}c_{i+2\hat{y},k\downarrow}+h.c\right)\rightarrow -t^{\prime}\sum_{i}\left(c_{ik\downarrow}^{\dagger}c_{i+2\hat{x},k\downarrow}-c_{ik\downarrow}^{\dagger}c_{i+2\hat{y},k\downarrow}+h.c\right).
 \end{eqnarray}
 For the $d_{xy}$ altermagnetic terms, the transformation proceeds analogously. As a consequence, the matrix representation for spin-down $c$ and $f$ fermions becomes the Hermitian conjugate of the corresponding spin-up matrix. This guarantees the absence of negative sign problem in the QMC simulations.

\section{II. Hybridization Function and Saddle Point}

The Kondo coupling is expressed as
\begin{eqnarray}
	J_K\sum_{l}\hat{\bm{S}}_l\cdot \hat{\bm{s}}_l,
\end{eqnarray}
where $\hat{\bm{S}}_l=\tfrac{1}{2}\sum_{\alpha\beta}\hat{f}_{l\alpha}^{\dagger}\bm{\sigma}_{\alpha\beta}\hat{f}_{l\beta}$ is the local spin operator in the pseudo-fermion representation, subject to the local constraint $\sum_{\alpha}\hat{f}_{l\alpha}^{\dagger}\hat{f}_{l\alpha}=1$.

Within the path-integral formalism of the multi-impurity Kondo model, the partition function takes the form
\begin{eqnarray}
	Z=\int D[c^{\dagger},c,f^{\dagger},f,\lambda]\,e^{-S},
\end{eqnarray}
with the action
\begin{eqnarray}
	S&=&\int_{0}^{\beta}d\tau \Bigg[\sum_{\bm{k}\sigma}c^{\dagger}_{\bm{k}\sigma}(\tau)(\partial_{\tau}+\epsilon_{\bm{k}\sigma})c_{\bm{k}\sigma}(\tau)+\sum_{l}f^{\dagger}_{l\sigma}(\tau)\left(\partial_{\tau}+ i\lambda_l\right)
    f_{l\sigma}(\tau)\Bigg]\nonumber\\
	&&-\int_{0}^{\beta}d\tau\,\frac{J_{K}}{2}\sum_{l}\left(\sum_{\beta}c_{l\beta}^{\dagger}(\tau)f_{l\beta}(\tau)\right)\left(\sum_{\alpha}f_{l\alpha}^{\dagger}(\tau)c_{l\alpha}(\tau)\right)-\sum_{l} i \lambda_l,
\end{eqnarray}
where $\lambda_l$ is introduced to enforce the single-occupancy the constraint.

By applying a HS transformation, we introduce the hybridization field $V_l$, leading to
\begin{eqnarray}
	-\frac{J_{K}}{2}\left(\sum_{\beta}c_{l\beta}^{\dagger}f_{l\beta}\right)\left(\sum_{\alpha}f_{l\alpha}^{\dagger}c_{l\alpha}\right)
	\;\;\rightarrow\;\;\frac{2\bar{V}_lV_l}{J_K}+\left(\sum_{\alpha}f_{l\alpha}^{\dagger}c_{l\alpha}\right)V_l+\bar{V}_l\left(\sum_{\alpha}c_{l\alpha}^{\dagger}f_{l\alpha}\right).
\end{eqnarray}

Accordingly, the action can be rewritten as
\begin{eqnarray}
	S[\bar{V},V,c^{\dagger},c,f^{\dagger},f,\lambda]=\int_{0}^{\beta}d\tau \left[\sum_{\sigma}\psi_{\sigma}^{\dagger}(\partial_{\tau}+h^{\sigma})\psi_{\sigma}+\sum_{j}\frac{2|V_j(\tau)|^2}{J}\right],
\end{eqnarray}
where
\begin{eqnarray}
	\psi_{\sigma}^{\dagger}=\big(c^{\dagger}_{\bm{k}_1\sigma},~\cdots,~c^{\dagger}_{\bm{k}_i\sigma},~\cdots,~c^{\dagger}_{\bm{k}_n\sigma},~f_{1\sigma}^{\dagger},~\cdots,~f_{j\sigma}^{\dagger},~\cdots,~f_{n_0\sigma}^{\dagger}\big),\nonumber
\end{eqnarray}
and
\begin{eqnarray}
	h_{\sigma}=\begin{pmatrix}
		E^{\sigma}&M^{\dagger}\\
		M&\Lambda
	\end{pmatrix},\qquad 
	M_{i,j}=\frac{V_{j}}{\sqrt{n}}e^{i\bm{k}_iR_j},\qquad 
	E^{\sigma}_{ij}=\delta_{ij}\epsilon_{\bm{k}_i\sigma},~\Lambda_{ij}=i\delta_{ij}\lambda_i.
\end{eqnarray}

\begin{figure}[h]
	\begin{center}
		\includegraphics[width=16cm]{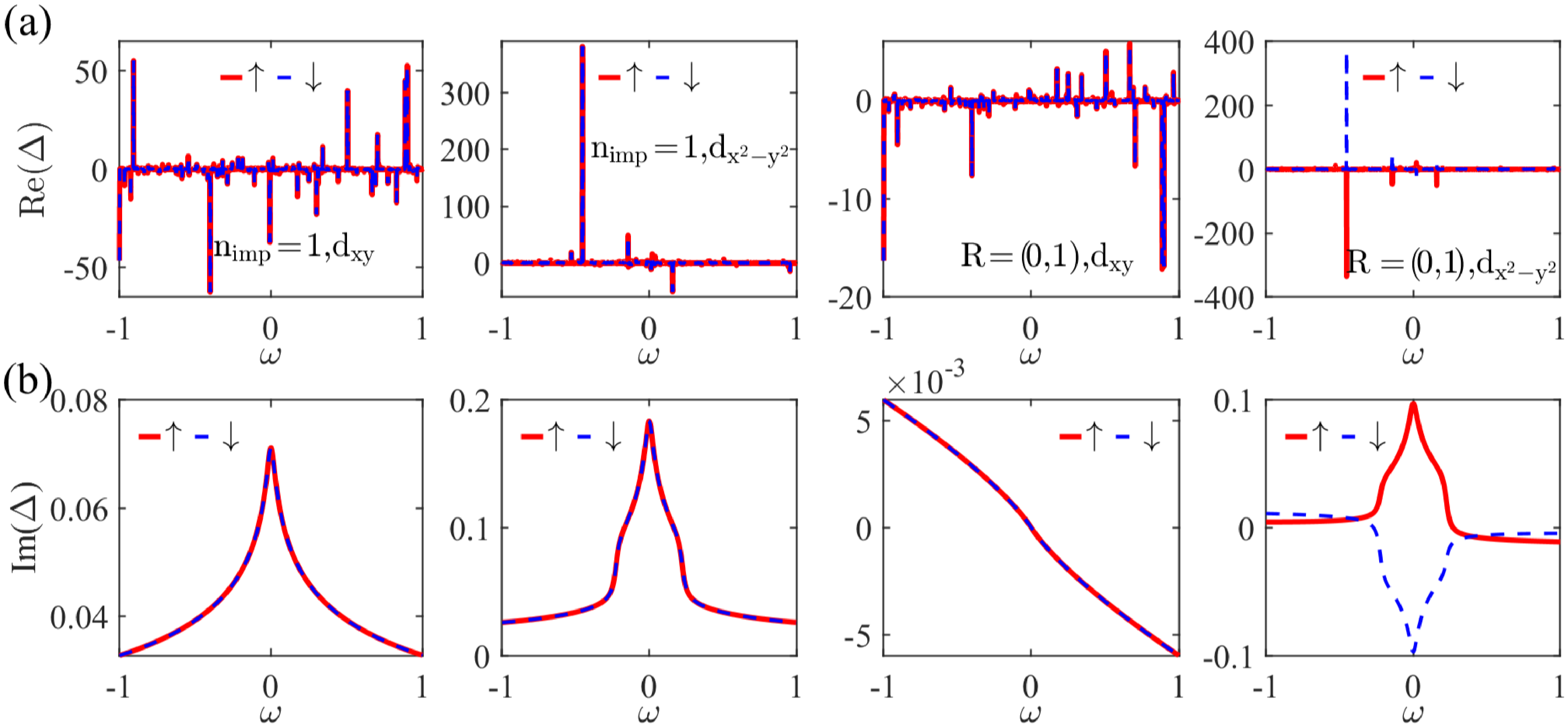}
	\end{center}
	\caption{
		(a) Real (Re) and (b) imaginary (Im) parts of the hybridization function $\Delta(\omega)$ for the single- and two-impurity Kondo models with altermagnetic terms, obtained with $t^{\prime}=0.4$ and $V_{l}=0.5$. The vector $\bm{R}=(0,1)$ denotes the spatial separation between the two impurities.}
	\label{figS1}
\end{figure}

By employing the Gaussian integral identity
\begin{eqnarray}
	\int \mathcal{D}(\bar{\phi},\phi)\,
	e^{-\bar{\phi}^{T}A\phi+\bar{v}^{T}\phi+\bar{\phi}^{T}v}
	= \det(A)\,e^{-\bar{v}A^{-1}v},
\end{eqnarray}
and assuming that the hybridization fields are static, we can integrate out the conduction electrons to obtain the effective action for the $f$ electrons. The partition function then becomes
\begin{eqnarray}
	Z=\int \mathcal{D}[\bar{V},V,f^{\dagger},f,\lambda]\,e^{-S_f},
\end{eqnarray}
where
\begin{eqnarray}
	S_{f}=\sum_{ln\sigma}f_{l\sigma}^{\dagger}(i\omega_n)(-i\omega_n+i\lambda_l)f_{l\sigma}(i\omega_n)
	+\sum_{njl\sigma}f_{j\sigma}^{\dagger}(i\omega_n)f_{l\sigma}(i\omega_n)\,\Delta_{jl}^{\sigma}(i\omega_n)
	+\sum_{l}\frac{2\bar{V}_lV_l}{J_{K}},
\end{eqnarray}
and the hybridization function is given by
\begin{eqnarray}
	\Delta_{jl}^{\sigma}(i\omega_n)
	=\sum_{\bm{k}}\frac{\bar{V}_jV_le^{i\bm{k}\cdot(\bm{R}_j-\bm{R}_l)}}{-i\omega_n+\epsilon_{\bm{k}\sigma}}.
\end{eqnarray}

For the single-impurity case, the hybridization function reduces to
\begin{eqnarray}
	\Delta^{\sigma}(i\omega_n)=\sum_{\bm{k}}\frac{|V|^2}{-i\omega_n+\epsilon_{\bm{k}\sigma}}.
\end{eqnarray}
Since the dispersion satisfies $\epsilon_{\bm{k}\sigma}\rightarrow \epsilon_{\bm{k}\bar{\sigma}}$ under the $C_{4z}$ rotation, the hybridization function obeys the relation $\Delta^{\sigma}(i\omega_n)=\Delta^{\bar{\sigma}}(i\omega_n)$, as illustrated in Fig.~\ref{figS1}(a)(b). It should be noted that the function is plotted after analytic continuation, $i\omega_n\rightarrow \omega+i\delta$. Consequently, the absence of spin splitting in the hybridization function enforces a spin-degenerate spectral function for both the local $f$ electrons and the conduction $c$ electrons.

For the two-impurity model, the diagonal components $\Delta_{11}^{\sigma}$ and $\Delta_{22}^{\sigma}$ are spin independent, whereas the off-diagonal components $\Delta_{12}^{\sigma}$ and $\Delta_{21}^{\sigma}$ acquire spin dependence. This spin asymmetry in the off-diagonal terms gives rise to spin splitting in the spectral functions of both the local $f$ electrons and the conduction $c$ electrons. For the case $\bm{R}=(0,1)$, the mirror symmetry $M_{yz}$ transforms the dispersion as $\epsilon_{\bm{k}\sigma}\rightarrow \epsilon_{\bm{k}\bar{\sigma}}$ in the $d_{xy}$ case, while leaving it invariant $\epsilon_{\bm{k}\sigma}\rightarrow \epsilon_{\bm{k}\sigma}$ in the $d_{x^2-y^2}$ case, resulting in spin degeneracy for the $d_{xy}$ altermagnetic symmetry, as shown in Fig.~\ref{figS1}(a)(b). Similarly, when $\bm{R}=(1,1)$, mirror symmetry along the diagonal direction ensures spin degeneracy for the $d_{x^2-y^2}$ case, whereas spin splitting persists in the $d_{xy}$ case.

The local Coulomb interaction of the $f$ electrons enforces the single-occupancy constraint but does not alter the functional form of the hybridization function.
 \begin{figure}[h]
 	\begin{center}
 		\includegraphics[width=16cm]{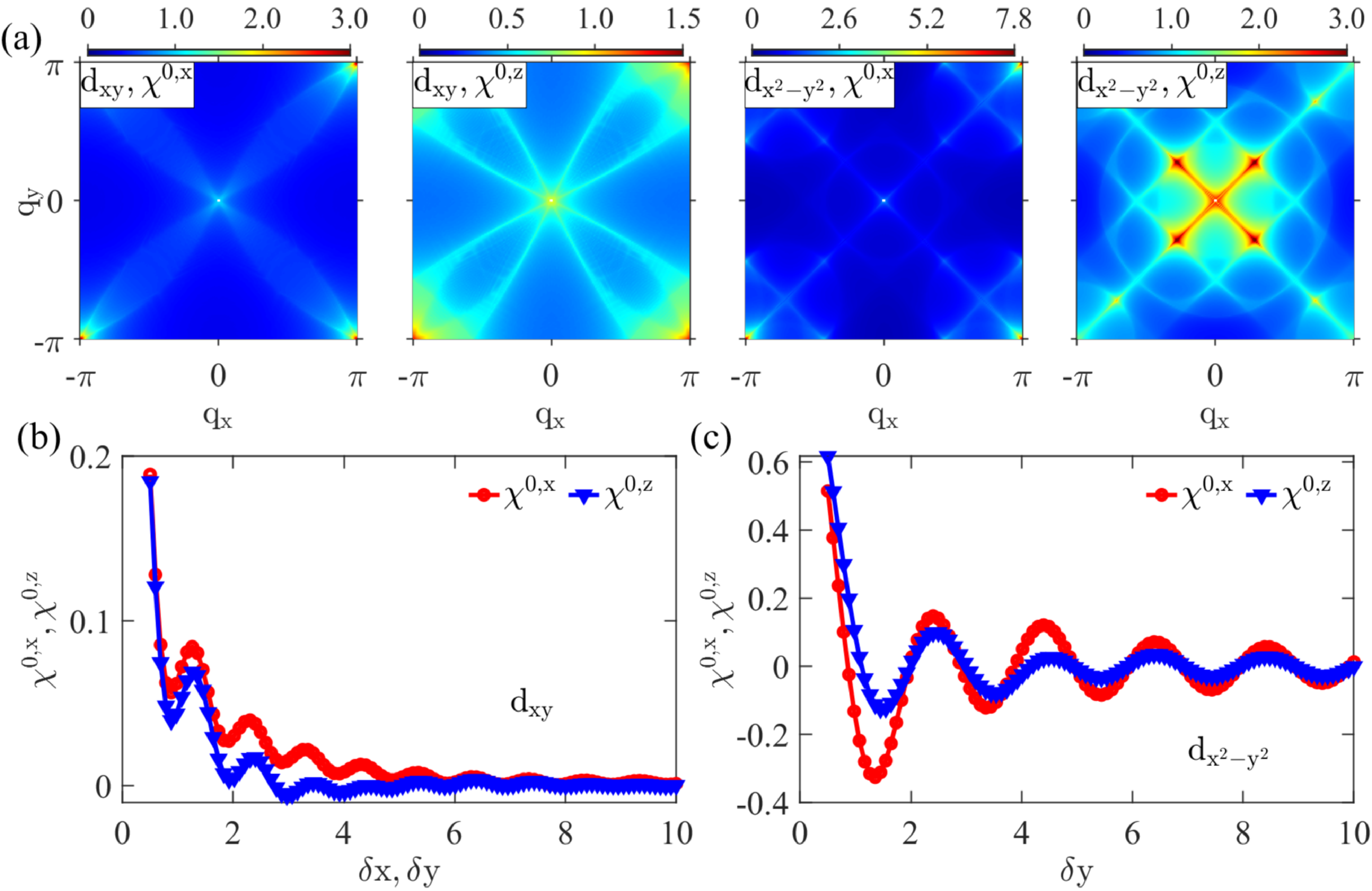}
 	\end{center}
 	\caption{(a) Momentum dependence of the static spin susceptibilities $\chi^{0,x}$ and $\chi^{0,z}$ for different altermagnetic terms, with $d_{x^2-y^2}$ and $d_{xy}$ indicating the altermagnetic symmetry. Real-space dependence of $\chi^{0,x}$ and $\chi^{0,z}$ along (b) the $y$ direction and (c) the diagonal direction.
 	}
 	\label{figS2}
 \end{figure}

 \section{III. RKKY Interaction}
 
 The two-impurity Kondo model can be expressed in momentum space as
 \begin{eqnarray}
 	H=\sum_{\bm{k}\sigma}\hat{c}_{\bm{k}\sigma}^{\dagger}\epsilon_{\bm{k}\sigma}\hat{c}_{\bm{k}\sigma}
 	+J_K\,\hat{\bm{S}}_{\bm{0}}\cdot \hat{\bm{s}}_{\bm{0}}
 	+J_K\,\hat{\bm{S}}_{\bm{R}}\cdot \hat{\bm{s}}_{\bm{R}},
 \end{eqnarray}
 where the dispersion relation is given by
 \begin{eqnarray}
 	\epsilon_{\bm{k}\sigma}=
 	\begin{cases}
 		-2t(\cos k_x+\cos k_y)-2t^{\prime}{\rm sgn}(\sigma)(\cos(k_x+k_y)-\cos(k_x-k_y)), & \text{for $d_{xy}$ altermagnetic case}, \\[6pt]
 		-2t(\cos k_x+\cos k_y)-2t^{\prime}{\rm sgn}(\sigma)(\cos 2k_x-\cos 2k_y), & \text{for $d_{x^2-y^2}$ altermagnetic case}.
 	\end{cases}
 \end{eqnarray}
 Here, $\hat{\bm{s}}_i$ denotes the local spin-density operator for $c$ electrons,
 \begin{eqnarray}
 	\hat{\bm{s}}_i=\tfrac{1}{2}\sum_{\alpha\beta}\hat{c}_{i\alpha}^{\dagger}\bm{\sigma}_{\alpha\beta}\hat{c}_{i\beta}.
 \end{eqnarray}

By employing the path-integral formalism for spins and fermions and integrating out the fermionic degrees of freedom, the action within second-order perturbation theory can be written as
\begin{eqnarray}
	S = S_0(\bm{n}) - \frac{J_K^2s^2}{2}\int_{0}^{\beta} d\tau \int_{0}^{\beta} d\tau' 
	\sum_{\mu\mu'} n_{0}^{\mu}(\tau)\,C^{0,\mu}_{\bm{R}}(\tau-\tau')\,n_{\bm{R}}^{\mu}(\tau'),
\end{eqnarray}
where $\langle n|\bm{S}_i|n\rangle = s\bm{n}_i$, and 
\begin{eqnarray}
	C^{0,\mu}_{\bm{R}}(\tau) = \left\langle T_{\tau}\, s_{\bm{0}}^{\mu}(\tau)\,s_{\bm{R}}^{\mu}(0)\right\rangle_0.
\end{eqnarray}
Here, $S_0(\bm{n})$ denotes the non-interacting part of the action. Thus, the RKKY interaction between two local moments separated by distance $\bm{R}$ is directly proportional to the non-interacting spin correlation function $C_{\bm{R}}^{0}(\tau)$.

Transforming to momentum space, the spin susceptibility takes the form
\begin{eqnarray}
	\chi^{0,\mu}_{\bm{q}}(i\omega_n) = -\sum_{\bm{k}}\sum_{\alpha\beta}
	\sigma_{\alpha\beta}^{\mu}\sigma_{\beta\alpha}^{\mu}\,
	\frac{f(\epsilon_{\bm{k}+\bm{q},\alpha}) - f(\epsilon_{\bm{k},\beta})}
	{i\omega_n + \epsilon_{\bm{k}+\bm{q},\alpha} - \epsilon_{\bm{k},\beta}}.
\end{eqnarray}
In particular, the static spin susceptibility is given by
\begin{eqnarray}
	\chi^{0,x}_{\bm{q}} = \chi^{0,y}_{\bm{q}} 
	= -2\sum_{\bm{k}}
	\frac{f(\epsilon_{\bm{k}+\bm{q},\uparrow}) - f(\epsilon_{\bm{k},\downarrow})}
	{\epsilon_{\bm{k}+\bm{q},\uparrow} - \epsilon_{\bm{k},\downarrow}}, 
\end{eqnarray}
and
\begin{eqnarray}
	\chi^{0,z}_{\bm{q}} 
	= -\sum_{\bm{k}\sigma}
	\frac{f(\epsilon_{\bm{k}+\bm{q},\sigma}) - f(\epsilon_{\bm{k},\sigma})}
	{\epsilon_{\bm{k}+\bm{q},\sigma} - \epsilon_{\bm{k},\sigma}},
\end{eqnarray}
where we omit the zero-frequency label $0$.

\begin{figure}[tb]
	\begin{center}
		\includegraphics[width=16cm]{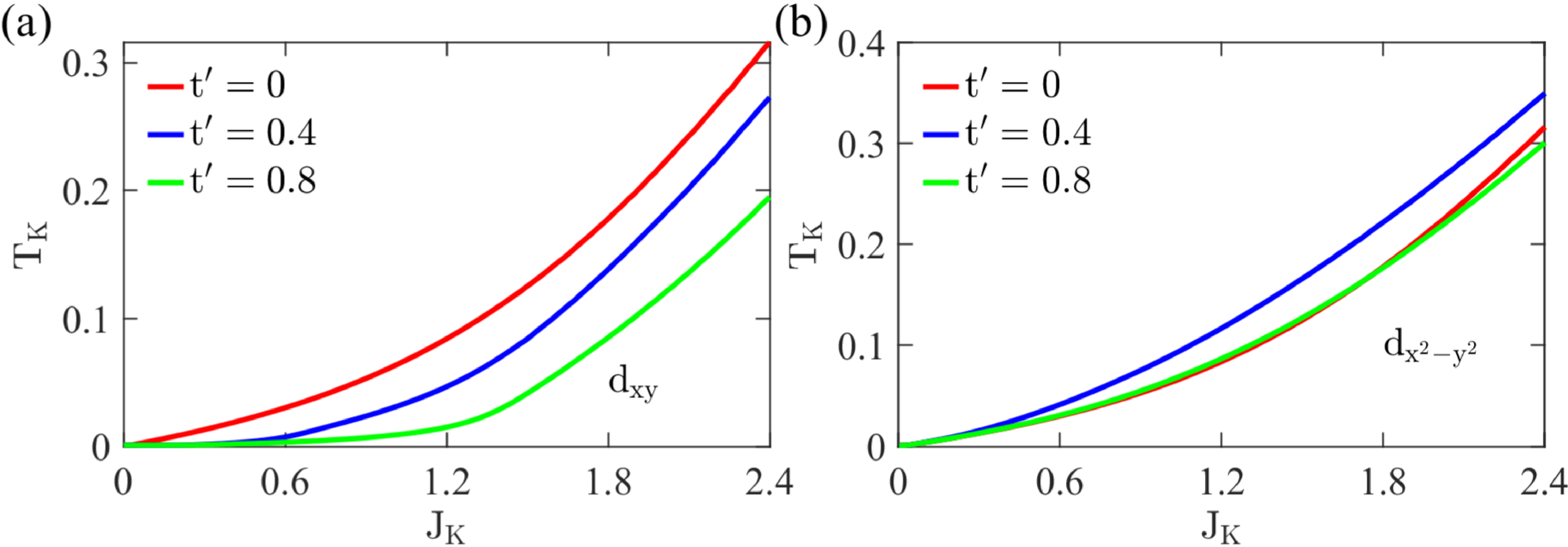}
	\end{center}
	\caption{ Mean-field Kondo temperature $T_{\rm K}$ as a function of the Kondo coupling $J_{\rm K}$ for different altermagnetic strengths $t^{\prime} = 0, 0.4, 0.8$, obtained from mean-field calculations for (a) the $d_{xy}$ case and (b) the $d_{x^2-y^2}$ case.
	}
	\label{figS3}
\end{figure}

In the absence of altermagnetic hopping ($t^{\prime}=0$), the dispersion satisfies 
$\xi_{\bm{k}\sigma} = \xi_{\bm{k}\bar{\sigma}}$ and $\xi_{\bm{k}+\bm{Q},\sigma} = -\xi_{\bm{k}\sigma}$, where $\bm{Q} = (\pi,\pi)$. Under these conditions, the spin susceptibilities become isotropic, 
\begin{eqnarray}
	\chi^{0,x} = \chi^{0,y} = \chi^{0,z},
\end{eqnarray}
and all components diverge at the wavevector $\bm{Q} = (\pi,\pi)$, reflecting the perfect nesting of the Fermi surface.

For both the $d_{xy}$ and $d_{x^2-y^2}$ cases, the dispersion satisfies 
$\xi_{\bm{k}+\bm{Q},\sigma}=-\xi_{\bm{k},\bar{\sigma}}$ with $\bm{Q}=(\pi,\pi)$. This nesting condition leads to a divergence in $\chi^{0,x}_{\bm{Q}}$, as illustrated in Fig.~\ref{figS2}(a). Consequently, $\chi^{0,x}_{\bm{Q}}>\chi_{\bm{Q}}^{0,z}$, in contrast to the isotropic case where $\chi^{0,x}_{\bm{Q}}=\chi^{0,z}_{\bm{Q}}$, providing a clear fingerprint of the altermagnetic contributions in the RKKY interaction. In real space, this manifests as oscillations along the $y$ direction, as shown in Fig.~\ref{figS2}(c). Along the diagonal direction, although oscillations persist, the values remain predominantly positive, as shown in Fig.~\ref{figS2}(b).

If the altermagnetic terms takes the form $-2t^{\prime}\,{\rm sgn}(\sigma)(\cos k_x - \cos k_y)$, the dispersion satisfies $\xi_{\bm{k}+\bm{Q},\sigma} = -\xi_{\bm{k},\sigma}$, leading to a divergence in $\chi^{0,z}_{\bm{Q}}$.

We note that, using Wicks theorem to compute the spin susceptibility of the  AM band structure, one can apply the result of the End Matter discussing the symmetry of the Green function, to show  that 
along a nodal line,
\begin{equation}
\chi^{0,x}_{\bm{R}}(i\omega_m) = \chi^{0,y}_{\bm{R}}(i\omega_m) = \chi^{z}_{\bm{R}}(i \omega_m).
\end{equation}

\section{IV. Mean-Field Calculations and Results}

\begin{figure}[tb]
	\begin{center}
		\includegraphics[width=16cm]{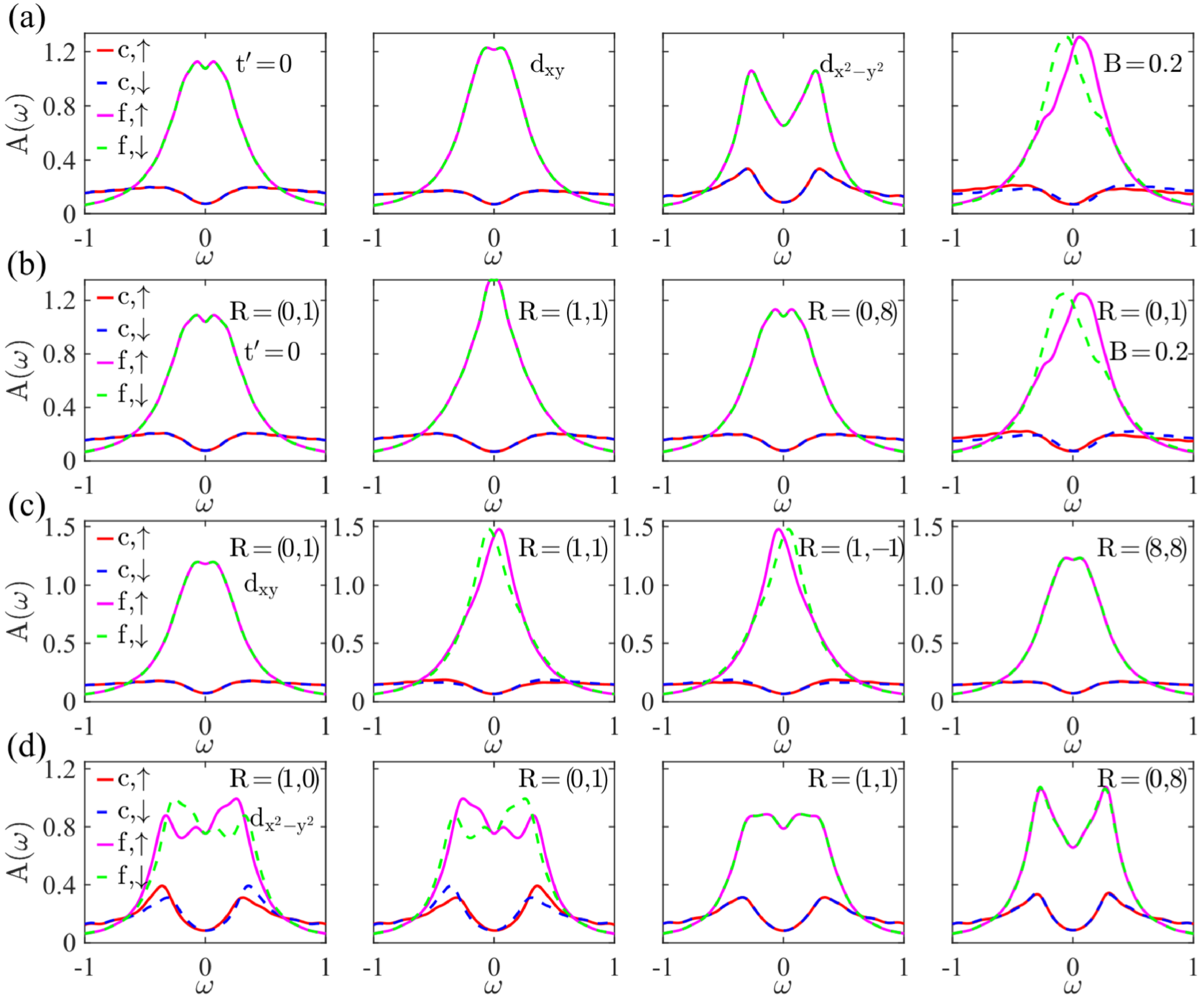}
	\end{center}
	\caption{ Spectral functions from the mean-field approximation. (a) Spin-dependent spectral functions of the $c$ and $f$ electrons for the single-impurity Kondo model under different conditions, where $d_{xy}$
    and $d_{x^2-y^2}$ indicate the symmetry of the altermagnetism, and $B=0.2$ denotes the applied magnetic field strength.  
		(b) Spectral functions for the two-impurity Kondo model for various impurity configurations, where $\bm{R}$ denotes the distance between the two impurities.  
		(c,d) Spin-dependent spectral functions for the two-impurity Kondo model with $d_{xy}$ and $d_{x^2-y^2}$ altermagnetic terms in the conduction-electron sector, respectively.
	}
	\label{figS4}
\end{figure}

For the multi-impurity Kondo model, the mean-field Hamiltonian can be expressed as
\begin{eqnarray}
	H_{\rm MF} = \sum_{\sigma} \hat{\psi}_{\sigma}^{\dagger} h^{\sigma} \hat{\psi}_{\sigma} + \mathrm{const},
\end{eqnarray} 
where
\begin{eqnarray}
	\hat{\psi}_{\sigma}^{\dagger} = \left( \cdots, \hat{c}_{\bm{k}\sigma}^{\dagger}, \cdots, \hat{f}_{1\sigma}^{\dagger}, \cdots, \hat{f}_{n_0\sigma}^{\dagger} \right), \quad
	h^{\sigma} = 
	\begin{pmatrix}
		E^{\sigma} & M^{\dagger} \\
		M & 0
	\end{pmatrix}, \quad
	E_{ij}^{\sigma} = \epsilon_{\bm{k}_i\sigma} \delta_{ij}, \quad
	M_{ij} = \frac{V_j}{\sqrt{n}} e^{i \bm{k}_i \cdot \bm{R}_j}.
\end{eqnarray}	
Here, we have set the Lagrange multiplier to zero. Owing to particle-hole symmetry this sets the  average occupancy of the $f$-electrons  to 
unity. 

Within the mean-field approximation, the Hamiltonian can be diagonalized as
\begin{eqnarray}
	\hat{\psi}_{\sigma}^{\dagger} h^{\sigma} \hat{\psi}_{\sigma} = \hat{\Gamma}_{\sigma}^{\dagger} \Lambda^{\sigma} \hat{\Gamma}_{\sigma}, \quad
	\hat{\psi}_{\sigma} = V_{\sigma} \hat{\Gamma}_{\sigma}, \quad
	V_{\sigma}^{\dagger} h^{\sigma} V_{\sigma} = \Lambda^{\sigma}.
\end{eqnarray}
The self-consistent hybridization field is determined by
\begin{eqnarray}
	V_l &=& -\frac{J_K}{2 \sqrt{n}} \sum_{\bm{k}\sigma} e^{-i \bm{k} \cdot \bm{R}_l} \langle \hat{c}_{\bm{k}\sigma}^{\dagger} \hat{f}_{\sigma} \rangle \nonumber \\
	&=& -\frac{J_K}{2 \sqrt{n}} \sum_{\bm{k}\sigma} e^{-i \bm{k} \cdot \bm{R}_l} \sum_{j} V_{\bm{k} j\sigma}^{*} V_{l j\sigma} f(E_{j}^{\sigma}),
\end{eqnarray}
where $f(E_j^{\sigma})$ is the Fermi-Dirac distribution function.

Within the mean-field framework, the Kondo transition temperature $T_K$ can be determined from the vanishing of the hybridization gap $V_l$, with the amplitude $|V_l|$ being uniform across different impurity sites. As illustrated in Fig.~\ref{figS3}(a) and (b), the Kondo transition temperature increases with the Kondo coupling $J_K$ for various altermagnetic strengths. The approximately linear behavior observed for the $d_{x^2-y^2}$ case may originate from the van Hove singularity at $\mu=0$ in the corresponding dispersion.

The Green function is defined as
\begin{eqnarray}
	G_{l\sigma}^{d}(\tau) = -\langle T_{\tau} \left( \hat{d}_{l\sigma}(\tau) \hat{d}_{l\sigma}^{\dagger}(0) \right) \rangle,
\end{eqnarray}
where $\hat{d}_{l\sigma} = \hat{c}_{l\sigma}$ 
or $\hat{f}_{l\sigma}$. In the imaginary frequency domain, the Green functions for the conduction and localized electrons are given by
\begin{eqnarray}
	G_{\sigma}^{c}(\bm{k},i\omega_n) = \sum_{j} \frac{V_{\bm{k} j\sigma} V_{\bm{k} j\sigma}^{*}}{i\omega_n - \Lambda_j^{\sigma}}, \qquad
	G_{l\sigma}^{f}(i\omega_n) = \sum_{j} \frac{V_{l+N,j\sigma} V_{l+N,j\sigma}^{*}}{i\omega_n - \Lambda_j^{\sigma}}.
\end{eqnarray}
The corresponding spectral functions can then be expressed as
\begin{eqnarray}
	A^{c}_{\sigma}(\omega) = \frac{\delta}{\pi} \sum_{\bm{k} j} \frac{V_{\bm{k} j\sigma} V_{\bm{k} j\sigma}^{*}}{(\omega - \Lambda_j^{\sigma})^2 + \delta^2}, \qquad
	A^{f}_{l\sigma}(\omega) = \frac{\delta}{\pi} \sum_{j} \frac{V_{l+N,j\sigma} V_{l+N,j\sigma}^{*}}{(\omega - \Lambda_j^{\sigma})^2 + \delta^2}.
\end{eqnarray}

If one adopts a real-space representation for $h^{\sigma}$, it takes the form
\begin{eqnarray}
	h^{\sigma} = 
	\begin{pmatrix}
		T^{\sigma} & Q^{\dagger} \\
		Q & 0
	\end{pmatrix}, \qquad T_{ij} = t_{ij}^{\sigma}, \quad Q_{ij} = V_j.
\end{eqnarray}
The local spectral functions of the $c$ and $f$ electrons are then given by
\begin{eqnarray}
	A^{f}_{l\sigma}(\omega) = \frac{\delta}{\pi} \sum_{j} \frac{V_{l+N,j\sigma} V_{l+N,j\sigma}^{*}}{(\omega - \Lambda_j^{\sigma})^2 + \delta^2}, \qquad
	A^{c}_{l\sigma}(\omega) =\frac{\delta}{\pi} \sum_{j} \frac{V_{lj\sigma} V_{lj\sigma}^{*}}{(\omega - \Lambda_j^{\sigma})^2 + \delta^2}.
\end{eqnarray}

\begin{figure}[tb]
	\begin{center}
		\includegraphics[width=16cm]{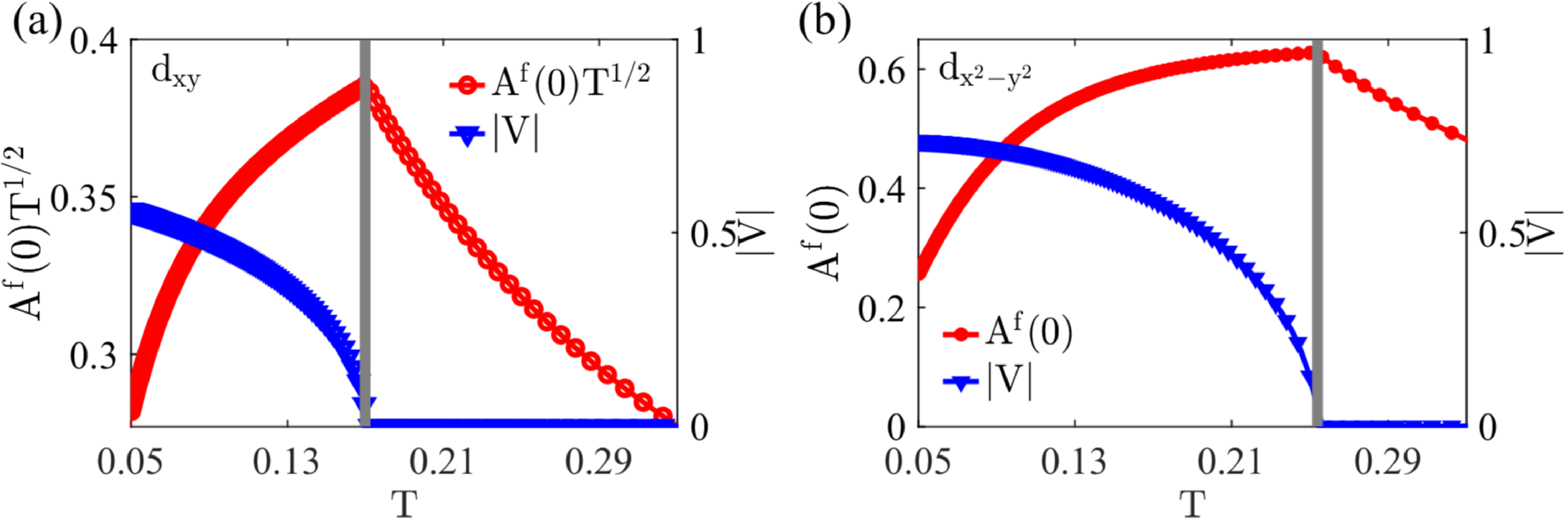}
	\end{center}
	\caption{ Temperature dependence of the hybridization gap and the zero-energy density of states $N^f(0)T^{1/2}$ and $N^f(0)$ for the $d_{xy}$ case with $\bm{R}=(1,1)$ and the $d_{x^2-y^2}$ case with $\bm{R}=(0,1)$. Parameters are $J_K = 2$ and $t^{\prime} = 0.4$.
	}
	\label{figS5}
\end{figure}

As shown in Fig.~\ref{figS4}, the spin-dependent spectral functions of the $c$ and $f$ electrons clearly illustrate the effects of the $d_{xy}$ and $d_{x^2-y^2}$ altermagnetic terms. For a single impurity, the altermagnetic terms do not lift the spin degeneracy, whereas an external magnetic field can induce splitting, as illustrated in Fig.~\ref{figS4}(a). In the two-impurity case with $t^{\prime}=0$, the spectral functions remain spin-degenerate regardless of the impurity separation, while the application of a magnetic field leads to a clear spin splitting, as shown in Fig.~\ref{figS4}(b). When the $d_{xy}$ or $d_{x^2-y^2}$ altermagnetic terms are present, the spin splitting of the spectral functions reflects the underlying altermagnetic symmetry, as shown in Fig.~\ref{figS4}(c) and (d). At sufficiently large impurity separations, the splitting disappears.

The zero-energy density of states can also be estimated within the mean-field framework using the relation 
\begin{eqnarray}
	A^f(0) \approx \frac{\beta}{\pi} G(\beta/2).
\end{eqnarray}
We observe that, for $\bm{R}=(1,1)$ in the $d_{xy}$ case
$A^f(0) T^{1/2}$ exhibits a pronounced peak at the Kondo transition temperature $T_K$, as shown in Fig.~\ref{figS5}, though $A^f(0)$ only has a kink. In contrast, for $\bm{R}=(0,1)$ in the  $d_{x^2-y^2}$ case,  
 the transition temperature $T_K$ can be identified by the peak in $A^f(0)$.

\section{V. Symmetry Analysis and Spectral Decomposition}

Under the combined operation of time-reversal and particle--hole transformations, the Green function
$\tilde{G}_{i\downarrow}(\tau)$ is mapped onto $\tilde{G}_{i\uparrow}(\beta-\tau)$.
This relation can be demonstrated explicitly as follows.
We start from the definition of the imaginary-time Green function for composite fermion,
\begin{eqnarray}
	\tilde{G}_{j\sigma}(\tau)
	= \langle T_{\tau}\,\hat{\psi}_{j\sigma}(\tau)\hat{\psi}_{j\sigma}^{\dagger}(0)\rangle .
\end{eqnarray}
For $\tau>0$, inserting a complete set of energy eigenstates yields the Lehmann representation \cite{Coleman2015},
\begin{eqnarray}
	\tilde{G}_{j\sigma}(\tau)
	= \sum_{n,m} e^{-\beta E_n} e^{(E_n - E_m)\tau}
	\langle n| \hat{\psi}_{j\sigma} |m\rangle
	\langle m| \hat{\psi}_{j\sigma}^{\dagger} |n\rangle .
\end{eqnarray}
Correspondingly, the associated spectral function can be expressed as
\begin{eqnarray}
	\hat{N}_{j\sigma}(\omega)
	= \sum_{n,m} \left(e^{-\beta E_n} + e^{-\beta E_m}\right)
	\langle n| \hat{\psi}_{j\sigma} |m\rangle
	\langle m| \hat{\psi}_{j\sigma}^{\dagger} |n\rangle
	\,\delta(\omega + E_n - E_m) .
\end{eqnarray}

Under the combined operation of time-reversal and particle-hole transformations,
the fermionic operator transforms as $\hat{\psi}_{j\downarrow} \;\rightarrow\; (-1)^j \hat{\psi}_{j\uparrow}^{\dagger}$. Accordingly, the matrix elements transform as $\langle n | \hat{\psi}_{j\downarrow} | m \rangle
	\;\rightarrow\;
	(-1)^j \langle n' | \hat{\psi}_{j\uparrow}^{\dagger} | m' \rangle$, 
where $|n'\rangle$ and $|m'\rangle$ denote the transformed many-body eigenstates.
As a consequence, the spectral functions satisfy the symmetry relation
\begin{equation}
	\tilde{N}_{j\downarrow}(\omega)
	=
	\tilde{N}_{j\uparrow}(-\omega).
\end{equation}

This symmetry immediately implies a corresponding relation for the imaginary-time
Green functions. Using the spectral representation
\begin{equation}
	\tilde{G}_{j,\sigma}(\tau)
	=
	\int d\omega\,
	\frac{\tilde{N}_{j,\sigma}(\omega)\,e^{-\omega\tau}}
	{e^{-\beta\omega}+1},
\end{equation}
we obtain
\begin{eqnarray}
	\tilde{G}_{j,\uparrow}(\beta-\tau)
	&=&
	\int d\omega\,
	\frac{\tilde{N}_{j,\uparrow}(\omega)\,e^{-\omega(\beta-\tau)}}
	{e^{-\beta\omega}+1}
	=
	\int d\omega\,
	\frac{\tilde{N}_{j,\uparrow}(\omega)\,e^{\omega\tau}}
	{e^{\beta\omega}+1}
	\nonumber\\
	&=&
	\int d\omega'\,
	\frac{\tilde{N}_{j,\uparrow}(-\omega')\,e^{-\omega'\tau}}
	{e^{-\beta\omega'}+1}
	=
	\int d\omega'\,
	\frac{\tilde{N}_{j,\downarrow}(\omega')\,e^{-\omega'\tau}}
	{e^{-\beta\omega'}+1}
	\nonumber\\
	&=&
	\tilde{G}_{j,\downarrow}(\tau),
\end{eqnarray}
which means that $\tilde{G}_{j,\uparrow}(\beta/2)=\tilde{G}_{j,\downarrow}(\beta/2)$.

We then obtain
\begin{eqnarray}
	\tilde{G}_{j,\uparrow}(\omega=0)
	&=&
	\int_{0}^{\beta} d\tau\, \tilde{G}_{j,\uparrow}(\tau)
	=
	\int_{0}^{\beta} d\tau\, \tilde{G}_{j,\downarrow}(\beta-\tau)
	\nonumber\\
	&=&
	\int_{0}^{\beta} d\tau'\, \tilde{G}_{j,\downarrow}(\tau')
	=
	\tilde{G}_{j,\downarrow}(\omega=0) .
\end{eqnarray}

Consequently, the imaginary-time integrated Green functions for the two spin
components are identical. In particular, the integral of the difference between
$\tilde{G}_{\bm{R},\uparrow}(\tau)$ and $\tilde{G}_{\bm{R},\downarrow}(\tau)$ over
imaginary time vanishes,
\begin{equation}
	\int_{0}^{\beta} d\tau\,
	\bigl[\tilde{G}_{\bm{R},\uparrow}(\tau)-\tilde{G}_{\bm{R},\downarrow}(\tau)\bigr]
	=
	\tilde{G}_{\bm{R},\uparrow}(\omega=0)
	-
	\tilde{G}_{\bm{R},\downarrow}(\omega=0)
	=
	0 .
\end{equation}

Instead, we introduce the following quantity,
\begin{eqnarray}
	\Delta \tilde{G}_{\bm{R}}
	&=&
	\left(
	\int_{0}^{\beta/2} d\tau
	-
	\int_{\beta/2}^{\beta} d\tau
	\right)
	\left[
	\tilde{G}_{\bm{R},\uparrow}(\tau)
	-
	\tilde{G}_{\bm{R},\downarrow}(\tau)
	\right]
	\nonumber\\
	&=&
	\left(
	\int_{0}^{\beta/2} d\tau
	-
	\int_{\beta/2}^{\beta} d\tau
	\right)
	\int d\omega\,
	\frac{e^{-\omega\tau}}
	{e^{-\beta\omega}+1}
	\left[
	\tilde{N}_{\bm{R},\uparrow}(\omega)
	-
	\tilde{N}_{\bm{R},\downarrow}(\omega)
	\right]
	\nonumber\\
	&=&
	\int d\omega\,
	\frac{1}{\omega}
	\frac{\left(1-e^{\beta\omega/2}\right)^2}
	{e^{\beta\omega}+1}
	\left[
	\tilde{N}_{\bm{R},\uparrow}(\omega)
	-
	\tilde{N}_{\bm{R},\downarrow}(\omega)
	\right]
	\nonumber\\
	&=&
	\int d\omega\,
	\left[
	\tilde{N}_{\bm{R},\uparrow}(\omega)
	-
	\tilde{N}_{\bm{R},\downarrow}(\omega)
	\right]
	f_{\beta}(\omega),
\end{eqnarray}
where $f_{\beta}(\omega)
	=
	\frac{1}{\omega}
	\left(
	1-\frac{1}{\cosh(\beta\omega/2)}
	\right)$
is an odd function of frequency.

\begin{figure}[tbh]
	\begin{center}
		\includegraphics[width=16cm]{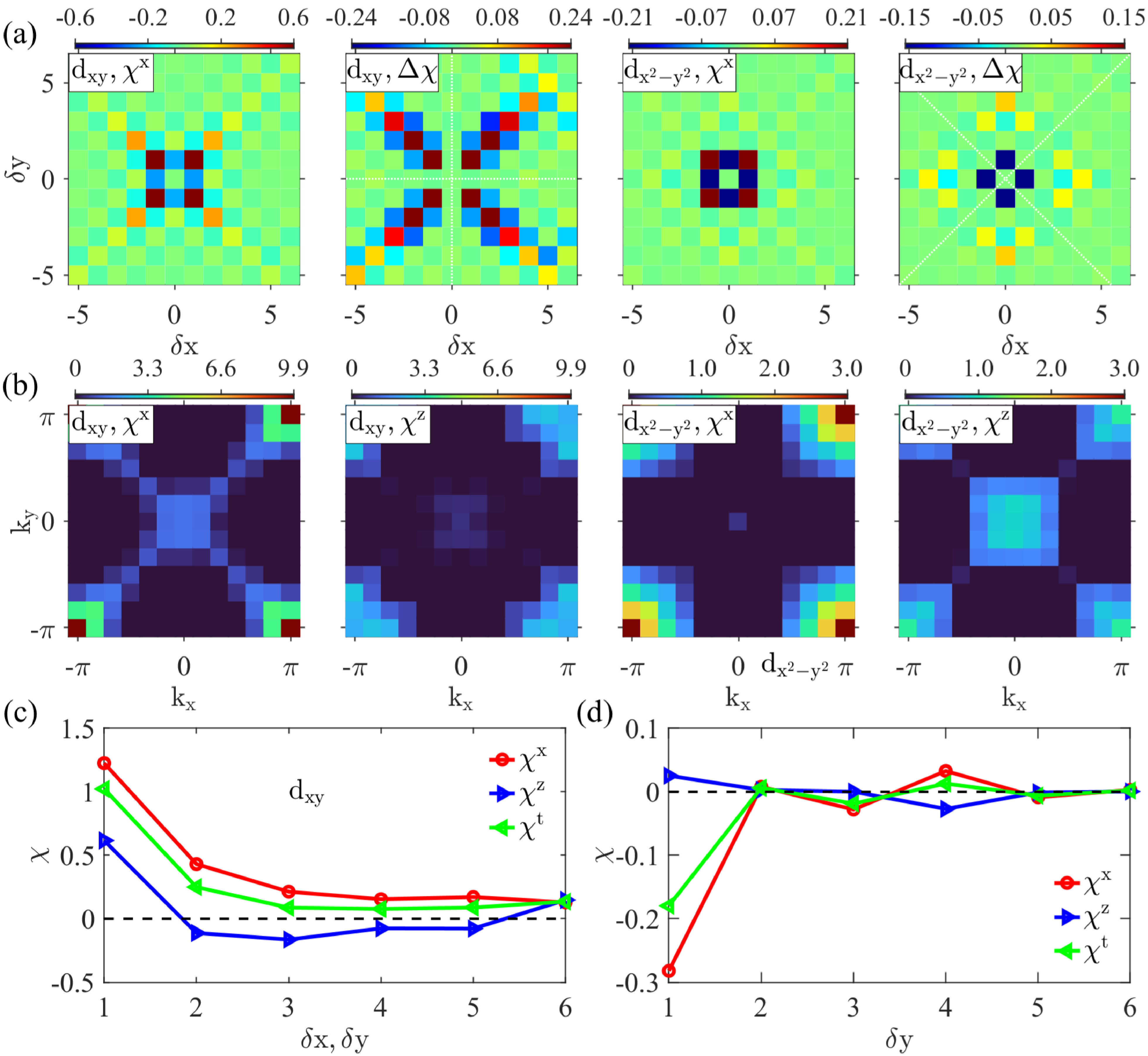}
	\end{center}
	\caption{ Spin susceptibility from QMC calculations. (a) Real-space distributions of the $x$-component static spin susceptibility
$\chi^{x}_{\bm{R}} $ and the susceptibility difference
$\Delta\chi = \chi^{x}_{\bm{R}} - \chi^{z}_{\bm{R}}$ for the $d_{xy}$ and $d_{x^2-y^2}$
altermagnetic cases at $T = 0.10$.
(b) Momentum-space distributions of the $x$- and $z$-component static spin
susceptibilities, $\chi^{x}_{\bm{k}}$ and $\chi^{z}_{\bm{k}}$, at $T = 0.10$.
(c) Distance dependence of the static susceptibilities $\chi^{x}_{\bm{R}} $, $\chi^{z}_{\bm{R}} $,
and the total susceptibility $\chi^{t}_{\bm{R}} = (2\chi^{x}_{\bm{R}}  + \chi^{z}_{\bm{R}})/3$ along the
diagonal direction for the $d_{xy}$ case at $T = 0.05$.
(d) Distance dependence of the static susceptibilities along the $y$ direction
for the $d_{x^2-y^2}$ case at $T = 0.05$.
	}
	\label{figS6}
\end{figure}

Consequently, $\Delta \tilde{G}_{\bm{R}}$ vanishes only if the spin-resolved spectral
difference
$\tilde{N}_{\bm{R},\uparrow}(\omega)-\tilde{N}_{\bm{R},\downarrow}(\omega)$
contains solely an even-frequency component.
In the present case, this difference is purely odd in frequency, leading to a finite
$\Delta \tilde{G}_{\bm{R}}$.

For the spin correlation function $C(\tau)$, a similar spectral decomposition can be carried out,
\begin{eqnarray}
	C(\tau)
	=
	\int d\omega\,
	\frac{\chi^{\prime\prime}(\omega)\,e^{-\omega\tau}}
	{1-e^{-\beta\omega}},
\end{eqnarray}
where $\chi^{\prime\prime}(\omega)$ denotes the imaginary part of the dynamical spin
susceptibility.

Accordingly, its imaginary-time integral,
\begin{equation}
	\chi(\omega=0)
	=
	\int_{0}^{\beta} d\tau\, C(\tau),
\end{equation}
can be analyzed in an analogous manner. The corresponding quantities can be expressed as
\begin{eqnarray}
	\chi(\omega=0)
=
	\int d\omega\,
	\frac{\chi^{\prime\prime}(\omega)}{\omega},~~~~
	C(\beta/2)
	=
	\int d\omega\,
	\frac{\chi^{\prime\prime}(\omega)}
	{2\sinh(\beta\omega/2)},~~~~
	C(0^{+})
	=
	\int d\omega\,
	\frac{\chi^{\prime\prime}(\omega)\,e^{-\omega 0^{+}}}
	{1-e^{-\beta\omega}} .
\end{eqnarray}
Here, all three quantities are predominantly determined by the low-energy behavior
of $\chi^{\prime\prime}(\omega)$.

In addition, $C(0^{+})$ represents an equal-weight superposition of contributions from
all imaginary-frequency components. By contrast, $\chi(\omega=0)$ isolates the
zero-frequency (static) component, and therefore more directly reflects the
low-energy spin physics of the system.

\section{VI. Static Spin Susceptibility}

In addition to the equal-time spin correlations obtained from quantum Monte Carlo simulations, we also analyze the static spin susceptibility, defined as
\begin{eqnarray}
\chi^{x(z)}_{\bm{R}} 
	= \int_0^\beta d\tau \,C_{\bm{R}}^{x(z)}(\tau),~~
C_{\bm{R}}^{x(z)}(\tau)=\langle T_{\tau}\, \hat{S}_{\bm{0}}^{x(z)}(\tau)\, \hat{S}_{\bm{R}}^{x(z)} \rangle-\langle \hat{S}_{\bm{0}}^{x(z)}\rangle\langle \hat{S}_{\bm{R}}^{x(z)}\rangle, 
	.\end{eqnarray}
Note that the total spin susceptibility satisfies the relation $\chi^{t}_{\bm{R}} =( 2\chi^{x}_{\bm{R}}  + \chi^{z}_{\bm{R}})/3$.

Similar to the equal-time spin correlation, the susceptibility difference
$\Delta \chi = \chi^{x}_{\bm{R}}  - \chi^{z}_{\bm{R}}$ exhibits nodal lines with the same symmetry:
along the $x$ and $y$ directions for the $d_{xy}$ case, and along the diagonal
directions for the $d_{x^2-y^2}$ case, as shown in Fig.~\ref{figS6}(a). This behavior
indicates the restoration of SU(2) spin symmetry along the nodal lines for the
corresponding altermagnetic states. In addition, for both cases, the transverse
susceptibility $\chi^{x}_{\bm{R}}$ displays a pronounced peak at $(\pi,\pi)$, as shown in
Fig.~\ref{figS6}(b).

Along the diagonal direction in the $d_{xy}$ case, as shown in Fig.~\ref{figS6}(c),
$\chi^{x}_{\bm{R}}$ remains positive for all impurity separations, indicating a
ferromagnetic interaction between the two impurities along the the diagonal direction. In contrast,
$\chi^{z}_{\bm{R}}$ becomes negative at certain distances, signaling the emergence of
antiferromagnetic correlations.

Along the $y$ direction in the $d_{x^2-y^2}$ case, both $\chi^{x}_{\bm{R}}$ and $\chi^{z}_{\bm{R}}$
exhibit oscillatory behavior, as shown in Fig.~\ref{figS6}(d).

\begin{figure}[tb]
	\begin{center}
		\includegraphics[width=16cm]{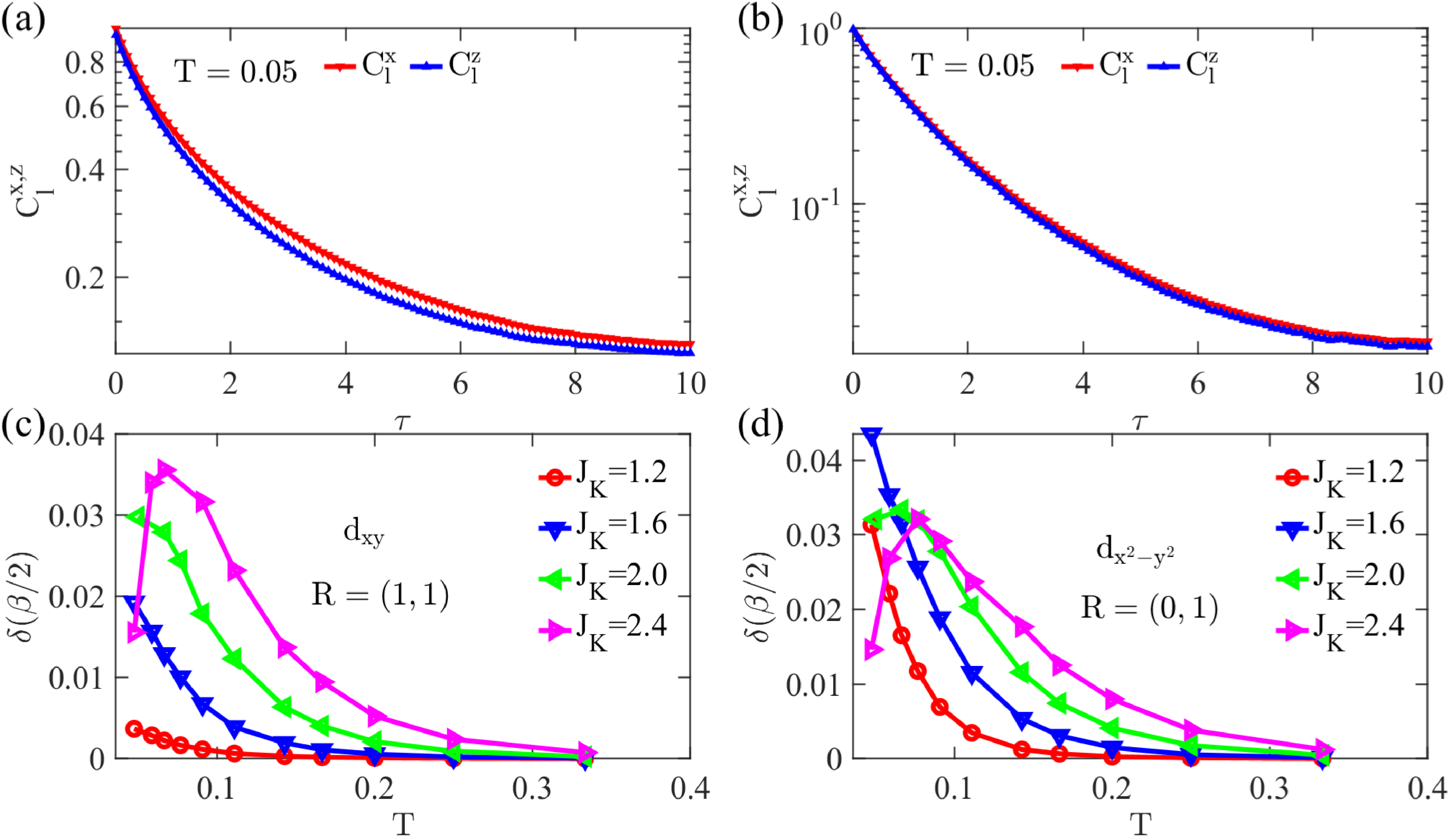}
	\end{center}
	\caption{ 
Anisotropic Kondo Screening for $d_{xy}$ case (a)(c) with $\bm{R}=(1,1)$ and $d_{x^2-y^2}$ case (b)(d) with $\bm{R}=(0,1)$. (a)(b) Local spin correlation functions $C_l^{x(z)}(\tau)$ as functions of imaginary
time $\tau$ at $T = 0.20$ and $J_K = 2.0$ for $d_{xy}$ and $d_{x^2-y^2}$ symmetry, respectively.
(c)(d) Anisotropy of Kondo screening,
$\delta(\beta/2)=\bigl[C_l^{x}(\beta/2) - C_l^{z}(\beta/2)\bigr] /
\bigl[C_l^{x}(\beta/2) + C_l^{z}(\beta/2)\bigr]$,
as a function of temperature at
$J_K = 1.2,\,1.6,\,2.0,\,2.4$ for $d_{xy}$ and $d_{x^2-y^2}$ altermagnetic symmetry.
}
	\label{figS7}
\end{figure}

\section{VII. Anisotropic Kondo Screening}

We have confirmed that Kondo screening persists even in the presence of altermagnetic terms in the End Matter; however, the screening is anisotropic.

The local spin correlation function $C_l^{x(z)}(\tau)$ is defined as
$C_l^{x(z)}(\tau) = \langle T_{\tau} \hat{S}_{\bm{0}}^{x(z)}(\tau) \hat{S}_{\bm{0}}^{x(z)} \rangle-\langle \hat{S}_{\bm{0}}^{x(z)}\rangle^2$. As shown in Fig. \ref{figS7}(a)(b), $C_l^{x}(\tau)$ at $T=0.05$ exhibits noticeable deviations from
$C_l^{z}(\tau)$ at large imaginary
times, corresponding to the low-energy regime. In contrast, at low $\tau$, the two correlation functions become nearly identical. This behavior signals anisotropic Kondo screening, with the
screening in the $z$ channel being more efficient than that in the $x$ channel.

Such anisotropy can be captured by the anistropic parameter $\delta(\tau)=\bigl[C_l^{x}(\tau) - C_l^{z}(\tau)\bigr] /
\bigl[C_l^{x}(\tau) + C_l^{z}(\tau)\bigr]$, and
suggests the emergence of an effective anisotropic Kondo
coupling $J_{K}^{x,z}$ in the low-energy theory, which would naturally appear in a
renormalization-group analysis. The magnitude of $\delta(\beta/2)$ in Fig.~\ref{figS7}(c)(d) increases upon lowering the
temperature, indicating that the anisotropy develops concomitantly with Kondo
screening in the presence of antiferromagnetic or ferromagnetic RKKY interactions. As $T \to 0$, the system approaches the fully Kondo-screened regime, accompanied by a reduction of $\delta(\beta/2)$.


\end{appendix}

\end{document}